\documentclass{aa}
\usepackage[utf8]{inputenc}
\usepackage[varg]{txfonts}
\usepackage{graphicx}
\usepackage{units}
\usepackage{siunitx}
\usepackage{longtable}
\usepackage{color}

\begin{document}

\newcommand{\correction}{}
\newcommand{\finalcorrection}{}

\title{Precise radial velocities of giant stars}
\subtitle{XI\correction{I}. Evidence against the proposed planet Aldebaran b\thanks{Based on observations collected at Lick Observatory, University of California.}}

\author{Katja Reichert\inst{1,2} \and Sabine Reffert\inst{1} \and Stephan Stock\inst{1} \and Trifon Trifonov\inst{3} \and Andreas Quirrenbach\inst{1}
}

\institute{Landessternwarte, Zentrum für Astronomie der Universität Heidelberg, Königstuhl 12, 69117 Heidelberg, Germany \and \correction{Astronomisches Rechen-Institut, Zentrum für Astronomie der Universität Heidelberg, Mönchhofstraße 12-14, 69120 Heidelberg, Germany} \and Max-Planck-Institut für Astronomie, Königstuhl 17, 69117 Heidelberg, Germany
}

\date{Received 6 August 2018 /
Accepted 1 March 2019}

\abstract
{Radial-velocity variations of the K giant star Aldebaran ($\alpha$~Tau) were first reported in the early 1990s. After subsequent analyses, the radial-velocity variability with a period of $\sim\!\!\unit[629]{d}$ has recently been interpreted as caused by a planet of several Jovian masses.}{We want to further investigate the hypothesis of an  extrasolar planet around Aldebaran.
}{We combine 165 new radial-velocity measurements from Lick Observatory with seven already published data sets comprising 373 radial-velocity measurements. We perform statistical analyses and investigate whether a Keplerian model properly fits the radial velocities. We also perform a dynamical stability analysis for a possible two-planet solution. \correction{
Furthermore, the possibility of oscillatory convective modes as cause for the observed radial-velocity
variability is discussed.}
}{
As best Keplerian fit to the combined radial-velocity data we obtain an orbit for the hypothetical planet with a smaller period ($P=\unit[607]{d}$) and a larger eccentricity ($e=0.33 \pm 0.04$) than the previously proposed one. However, the residual \correction{scatter around that fit is still large, with a standard deviation of $\unit[117]{ms^{-1}}$}.
In 2006/2007, the statistical power of the $\sim\!\!\unit[620]{d}$ period showed a temporary but significant decrease.
Plotting the growth of power in reverse chronological order reveals that a period around $\unit[620]{d}$ is clearly present in the newest data but not in the data taken before $\sim\!\!2006$. Furthermore, an apparent phase shift between radial-velocity data and orbital solution is observable at certain times. A two-planet Keplerian fit matches the data considerably better than a single-planet solution, but poses severe dynamical stability issues.
}{The radial-velocity data from Lick Observatory do not further support but in fact weaken the hypothesis of a substellar companion around Aldebaran. Oscillatory convective modes might be a plausible alternative explanation of the observed radial-velocity variations.}

\maketitle

\titlerunning{RV Variations of Aldebaran}
\authorrunning{K. Reichert et al.}

\begin{table}
\caption{ Stellar parameters of Aldebaran. The values for the mass, radius, $\log g$, $T_\mathrm{eff}$, luminosity, and age are given for both possibilities that Aldebaran is either on the RGB or the HB.}
\label{tab:stellar_parameters}
\centering
\begin{tabular}{l r@{}c@{}l r r}
\hline\hline \vspace{-0.3cm} \\
Parameter & {Value}& & & & Ref. \\
\hline \vspace{-0.3cm} \\
Spectral type & {K5 III} & & & & 1 \vspace{0.03cm} \\
Parallax $[\mathrm{mas}]$& 48.94 &$\pm$& 0.77 & &2 \vspace{0.03cm} \\
Distance $[\mathrm{pc}]$ & 20.43 &$\pm$& 0.32 & &2 \vspace{0.03cm} \\
$B-V$ $[\mathrm{mag}]$ & 1.538 &$\pm$& 0.008 & &3 \vspace{0.03cm} \\
$B_T$ $[\mathrm{mag}]$& 2.937 &$\pm$& 0.006 & &4 \vspace{0.03cm} \\
$V_T$ $[\mathrm{mag}]$& 1.160 &$\pm$& 0.011 & &4 \vspace{0.03cm} \\ 
$\mathrm{[Fe/H]}$ [dex] & --0.36 &$\pm$& 0.10  & &5 \vspace{0.03cm} \\
\hline \hline \vspace{-0.3cm} \\
Evol.\ stage: & RGB & & & HB & Ref.\\
Probability: & $97.8\%$ & & & $2.2\%$ & \\
\hline \vspace{-0.3cm} \\
Mass $[\mathrm{M}_\odot]$ & {$0.91^{+0.04}_{-0.02} $}& & &$0.96^{+0.04}_{-0.13}$ & 6 \vspace{0.03cm}\\
Radius $[\mathrm{R}_\odot]$ & {$44.01^{+0.74}_{-0.71}$}& & &$44.15^{+0.67}_{-0.92}$ & 6 \vspace{0.03cm}\\
log $g$ [cgs] & {$1.12^{+0.02}_{-0.02}$}& & &$1.12^{+0.03}_{-0.02}$ & 6 \vspace{0.03cm}\\
$T_{\mathrm{eff}}$ [K] & {$3901^{+10}_{-10}$}& & &$3899^{+11}_{-11}$ & 6 \vspace{0.03cm}\\
Luminosity $[\mathrm{L}_\odot]$ & {$402^{+11}_{-10}$}& & &$402^{+12}_{-13}$ & 6 \vspace{0.03cm}\\
Age [Gyr] & {$10.2^{+0.9}_{-0.9}$}& & &$9.8^{+0.9}_{-0.9}$ & 6 \vspace{0.03cm}\\
\hline
\end{tabular}
\tablebib{
(1)~\citet{Gray_2006}; (2) \citet{van_Leeuwen_2007}; (3) \citet{1997ESASP1200.....E}; (4) \citet{tycho};
(5) \citet{Hekker_2007}; (6)  \cite{Stock_2018}}
\end{table}

\section{Introduction}

The number of detected extrasolar planets has grown continuously since the early 1990s but has especially increased since 2014 due to the very successful \textit{Kepler} space mission. Most of the detected extrasolar planets orbit main-sequence (MS) stars. Only a small fraction of them have been found around giant stars; the first one (around $\iota$ Draconis) was discovered in 2002 by \cite{Frink_2002}. The number of detected planetary companions around giant stars has almost doubled within recent years. Meanwhile, 112 substellar companions\footnote{\url{https://www.lsw.uni-heidelberg.de/users/sreffert/giantplanets.html}} around 102 G and K giant stars have been found.

Due to the effects of the evolution of the host stars into giants on the orbits of their companions, such systems are of special interest as they allow us to get an idea about the dramatic destiny of our own solar system. Furthermore, evolved G or K giant stars are suitable targets for radial-velocity (RV) planet search programs due to their numerous absorption lines, compared to their F or A progenitor stars \citep{Quirrenbach_2011}. In addition, detecting planets around their fast-rotating progenitor stars on the MS using spectroscopic methods is usually very challenging, whereas the further evolved giant stars enable the search for extrasolar planets around stars in a higher stellar mass range.

\cite{Hatzes_1993} were the first to report low-amplitude variations with a period of 643 days in the RV data of the K giant star Aldebaran (HIP 21421, HD 29139), also known as $\alpha$~Tau. Even two years before the first detection of an extrasolar planet around a solar-type star, they already assumed that one explanation for these variations could be a planetary companion with a minimum mass of $11.4$ Jupiter masses (assuming a stellar mass of $\unit[2.5]{M_\odot}$), since periods in this range are much too long to be caused by radial pulsations. However, they could not exclude that the variations are caused by rotational modulation of surface features or non-radial pulsations.

Furthermore, they found these variations to be consistent in amplitude and phase for at least ten years by comparison with previous measurements, as part of their McDonald Observatory Planetary Search (MOPS) program, and measurements of \cite{Walker_1989}. In addition, there were no hints for a period of $\sim\!\!\unit[643]{d}$ in a subsequent analysis of the spectral line shapes by \cite{Hatzes_1998}. However, they found a period of 50 days in the line bisector measurements and assumed it to be due to stellar oscillations. Astrometric measurements of \cite{Gatewood_2008} could not refute the planet hypothesis of \cite{Hatzes_1998}, though \cite{Gatewood_2008} notes that the observed motion is ``marginal at best and smaller than suggested by \cite{Hatzes_1998}''. \cite{Gatewood_2008} also mentioned that the possible companion is likely to have a mass of 3 to 4 Jovian masses and not as large as the mass of $\unit[11]{M_\mathrm{Jup}}$ which \cite{Hatzes_1998} had derived.

\cite{Hatzes_2015} (hereafter H15) combined data taken between 1980 and 2013 at the Landessternwarte Tautenburg (TLS), the McDonald Observatory (McD), the Canada-France-Hawaii Telescope (CFHT), the Dominion Astrophysical Observatory (DAO), and at the Bohyunsan Optical Astronomy Observatory (BOAO), to re-investigate the nature of the long-period and long-lived RV variations in Aldebaran. They used three different data sets from McD, the original measurements using the $\unit[2.1]{m}$ telescope (hereafter ``McD-2.1m'' data set), measurements with the coude spectrograph in the so-called cs11 focus (hereafter ``McD-CS11'' data set), and measurements using the Tull Spectrograph (hereafter ``McD-Tull'' data set).

They were able to confirm the presence of a period in the range of $\sim\!\!\unit[629]{d}$ in the RV data but did not find similar periods in any activity indicator. Therefore, they concluded that these variations are not caused by stellar activity and assumed them to be due to a substellar companion.
Their orbital solution yielded a period $P=\unit[628.96 \pm 0.90]{d}$, an eccentricity $e=0.10 \pm 0.05,$ and a RV amplitude $K=\unit[142.1 \pm 7.2]{ms^{-1}}$ as parameters for the potential planetary companion. 

Although the RV data published by H15 already extend over a time span of almost 34 years we add available data from Lick Observatory to those published by H15, as the Lick data were partially taken during a time with sparse sampling in the H15 data sets, to re-examine the planet hypothesis.

We note that \cite{Farr_2018} have recently presented 302 new SONG RVs for Aldebaran obtained over a time span of about 120~days. They model their new RVs together with those from H15 with an orbiting planet plus a Gaussian process term due to acoustic oscillations. We show in Appendix~\ref{appa} that their modelling of the acoustic oscillations (with a dominating period around 6 days) does not help with the interpretation of the long-term RV variation (many hundreds of days). Their model  neither refutes nor confirms the planet hypothesis, although it does help with the asteroseismic determination of stellar parameters.

\begin{figure}
\resizebox{\hsize}{!}{\includegraphics{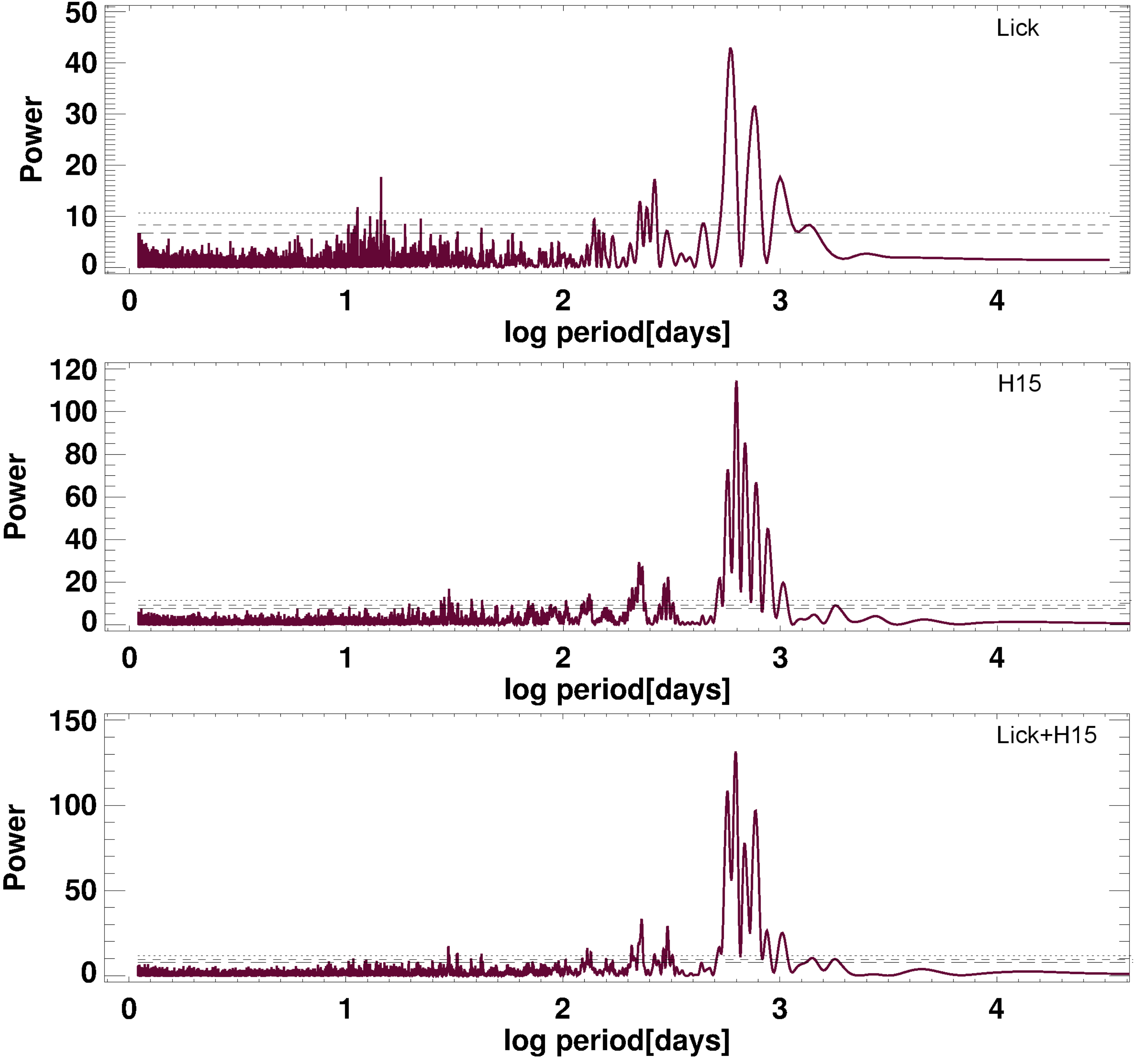}}
\caption{\textit{Top:} L-S periodogram for the Lick data. The most significant peak is at a period of $\unit[587.84]{d}$ with a FAP of $1.6 \cdot 10^{-17}$. \textit{Middle:} L-S periodogram for the H15 data. The most significant period is $\unit[629.08]{d}$ and has a FAP of less than $10^{-41}$. \textit{Bottom:} L-S periodogram for the Lick data combined with the data of H15. The most significant peak is found at a period of $\unit[625.02]{d}$ with a FAP of less than $10^{-41}$. In all three panels, the significance thresholds represented by the vertical dashed lines correspond to FAPs of $\unit[0.1]{\%}$, $\unit[1]{\%}$, and $\unit[5]{\%}$, respectively.}
\label{fig:Losca_Lick_Hatzes_Comb}
\end{figure}

\begin{table*}
\caption{Significant periods and their FAPs in the L-S periodogram of the Lick data (left) and Lick + H15 data (right). The periods are sorted by L-S power of the combined data in a descending order. It should be noted that the $\unit[263.35]{d}$ period is the fifth most significant period in the Lick data but was put in last place for comparison with the $\unit[264.69]{d}$ in the combined data sets. The significance threshold is a FAP of $\unit[0.1]{\%}$.}
\label{tab:List_of_Periods}
\centering
\begin{tabular}{c c c c c c}
\hline\hline \vspace{-0.3cm} \\
& Lick & & & Lick + H15 &\\
\hline \vspace{-0.3cm} \\
Period [d] & Power & FAP & Period [d] & Power & FAP\\
\hline \vspace{-0.3cm} \\
\ldots & \ldots & \ldots & 625.02 & 131.577 & $  < 10^{-41} $\\
587.84 & 42.962 & $ 1.6 \cdot 10^{-17} $ & 573.24 & 108.554 & $  < 10^{-41} $ \\
765.56 & 31.578 & $ 1.4 \cdot 10^{-12} $ & 775.02 & 96.692 & $ 2.2 \cdot 10^{-40} $ \\
\ldots & \ldots & \ldots & 687.08 & 77.801 & $ 3.6 \cdot 10^{-32} $ \\
\ldots & \ldots & \ldots & 230.66 & 33.182 & $ 8.6 \cdot 10^{-13} $ \\
\ldots & \ldots & \ldots & 303.69 & 29.083 & $ 5.2 \cdot 10^{-11} $ \\
\ldots & \ldots & \ldots & 872.77 & 26.320 & $ 8.2 \cdot 10^{-10} $ \\
14.44 & 17.644 & $ 1.6 \cdot 10^{-6}$ & \ldots & \ldots & \ldots \\
997.55 & 17.628 & $ 1.7 \cdot 10^{-6}$ & 1030.62 & 25.039 & $ 2.9 \cdot 10^{-9} $ \\
225.47 & 12.930 & $ 1.8 \cdot 10^{-4}$ & 223.74 & 18.998 & $ 1.2 \cdot 10^{-6} $ \\
11.25 & 11.776 & $ 5.6 \cdot 10^{-4}$ & \ldots & \ldots & \ldots \\
242.05 & 11.654 & $ 6.3 \cdot 10^{-4}$ & \ldots & \ldots & \ldots \\
\ldots & \ldots & \ldots & 207.89 & 17.752 & $4.3 \cdot 10^{-6} $ \\
\ldots & \ldots & \ldots & 29.68 & 17.212 & $7.4 \cdot 10^{-6} $ \\
\ldots & \ldots & \ldots & 29.55 & 16.774 & $1.1 \cdot 10^{-5} $ \\
\ldots & \ldots & \ldots & 526.51 & 16.686 & $1.2 \cdot 10^{-5} $ \\
\ldots & \ldots & \ldots & 129.52 & 16.029 & $2.4 \cdot 10^{-5} $ \\
\ldots & \ldots & \ldots & 290.92 & 15.549 & $3.8 \cdot 10^{-5} $ \\
\ldots & \ldots & \ldots & 134.55 & 13.621 & $2.6 \cdot 10^{-4} $ \\
\ldots & \ldots & \ldots & 32.65 & 13.107 &  $4.4 \cdot 10^{-4} $ \\
\ldots & \ldots & \ldots & 42.16 & 12.777 & $6.0 \cdot 10^{-4} $ \\
\ldots & \ldots & \ldots & 32.48 & 12.618 &  $7.0 \cdot 10^{-4} $ \\
\ldots & \ldots & \ldots & 29.82 & 12.571 &  $7.4 \cdot 10^{-4} $ \\
263.35 & 17.233 & $ 2.5 \cdot 10^{-6}$ & 264.69 & 12.265 &  $9.9 \cdot 10^{-4} $ \\
\ldots & \ldots & \ldots & 319.73 & 11.874 &  $1.4 \cdot 10^{-3} $ \\
\hline
\end{tabular}
\end{table*}

\section{Stellar parameters of Aldebaran}

\object{Aldebaran} is a K5 III giant star \citep{Gray_2006} and is located at a distance of about 20 pc \citep{van_Leeuwen_2007} in the constellation of Taurus. With an average apparent magnitude of $m_\mathrm{V}=0.87$ \citep{1997ESASP1200.....E} it is one of the brightest stars in the sky.

We determined the stellar parameters of Aldebaran using evolutionary tracks by \cite{Bressan_2012} and a Bayesian inference method based on the approach by \cite{Jorgensen_2005} with some improvements; see \cite{Stock_2018} for details. Our method is capable of providing a probability for each of the two degenerate post-MS evolutionary stages in the HRD. Additionally, we provide asymmetric 1\,$\sigma$ confidence intervals, as they describe the non-symmetrical probability density functions more properly. Our determined parameters are the mode values (maximum) of the probability density functions. The mode is less biased than the mean, especially if the distribution is bi-modal or truncated \citetext{see also \citealt{Jorgensen_2005}}. As input parameters we used the metallicity [Fe/H]$=-0.36\pm 0.10$ as given by \cite{Hekker_2007}, the Hipparcos parallax of $\unit[48.94\pm 0.77]{mas}$ \citep{van_Leeuwen_2007}, and $B$, $V$ photometry provided by the Hipparcos catalog \citep{1997ESASP1200.....E}.

\correction{Aldebaran is not included in the first or second Gaia data release.
Gaia DR2 however provides a parallax of $\unit[47.34 \pm 0.11]{mas}$ \citep{GaiaDR2} for a star in the close vicinity of Aldebaran, $\alpha$~Tau~B. It is a red dwarf with a mean magnitude of $\unit[11.9]{mag}$ in the G band \citep{GaiaDR2}. Its angular distance to Aldebaran of only about $\unit[650]{AU}$ as well as its proper motion, which is comparable to that of Aldebaran, suggest that the red dwarf is gravitationally bound to Aldebaran. It would thus make sense to adopt the Gaia parallax of the companion as the overall parallax of the system. However, since the parallax of Aldebaran was determined already very precisely in \cite{van_Leeuwen_2007},
we used this value for determining its stellar parameters.}  

To avoid any biases which can occur due to converting parallax measurements to absolute magnitudes, we used the astrometry-based luminosity \citep{Arenou_1999}, a quantity which is linear in the parallax. 

We determine that most probably ($P=\unit[97.8]{\%}$) Aldebaran is a very late RGB star with a current mass of $M_*=\unit[0.91^{+0.03}_{-0.02}]{M_\odot}$, which includes a Reimer's mass loss of $M_*=\unit[0.02]{M_\odot}$ calculated using $\eta=0.2$ as suggested by \cite{Miglio_2012}. For the radius we obtain $R_*=\unit[44.01^{+0.74}_{-0.71}]{R_\odot}$ and for the age $\tau=\unit[10.2^{+0.9}_{-0.9}]{Gyr}$. Table~\ref{tab:stellar_parameters} also lists the stellar parameter for the unlikely case that Aldebaran is already on the HB.

Our determined mass is smaller than that of H15, who determined $M_*=\unit[1.13 \pm 0.11]{M_\odot}$ as well as a radius of $R_*=\unit[36.68 \pm 2.46]{R_\odot}$ and an age of $\tau=\unit[6.6 \pm 2.4]{Gyr}$. The different stellar parameters are caused by the lower metallicity estimate of $\mathrm{[Fe/H]}=-0.36\pm0.1$ by \cite{Hekker_2007} compared to the estimate by H15  of $\mathrm{[Fe/H]}=-0.27 \pm 0.05$. \cite{Mozurkewich_2003} interferometrically measured a diameter of $\unit[21.10 \pm 0.21]{mas}$ for the limb-darkened case, corresponding to a radius of $\unit[46.3 \pm 0.9]{R_\odot}$, while \citet{Richichi_2005} determine $\unit[20.58\pm 0.03]{mas}$, corresponding to $\unit[45.2 \pm 0.7]{R_\odot}$. Our value for the radius of Aldebaran is thus in excellent agreement with the interferometrically determined values.

\begin{figure*}
\centering
\includegraphics[width=9cm, trim=1cm 0.5cm 0.5cm 0.9cm,clip]{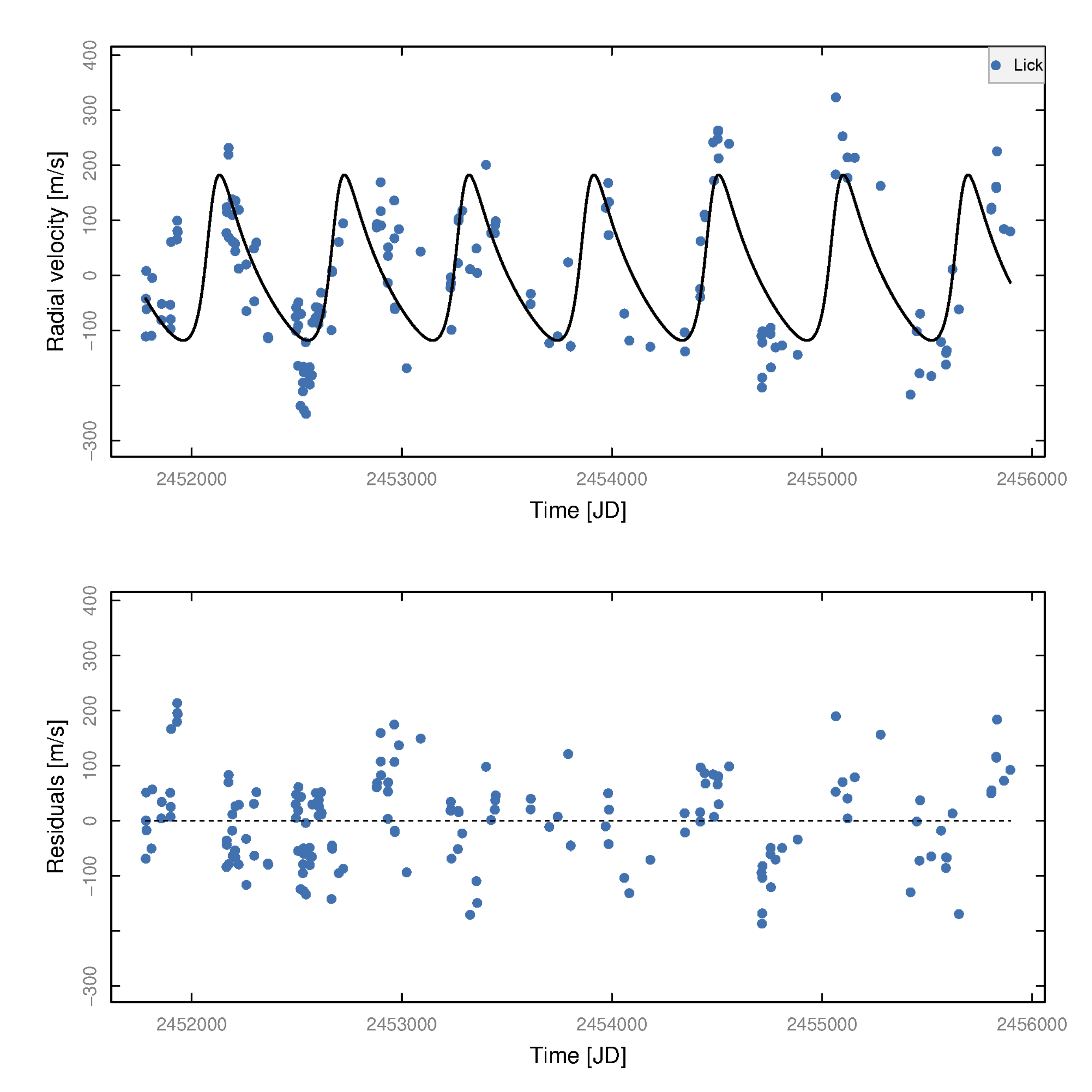}
\includegraphics[width=9cm, trim=1cm 0.5cm 0.5cm 0.9cm,clip]{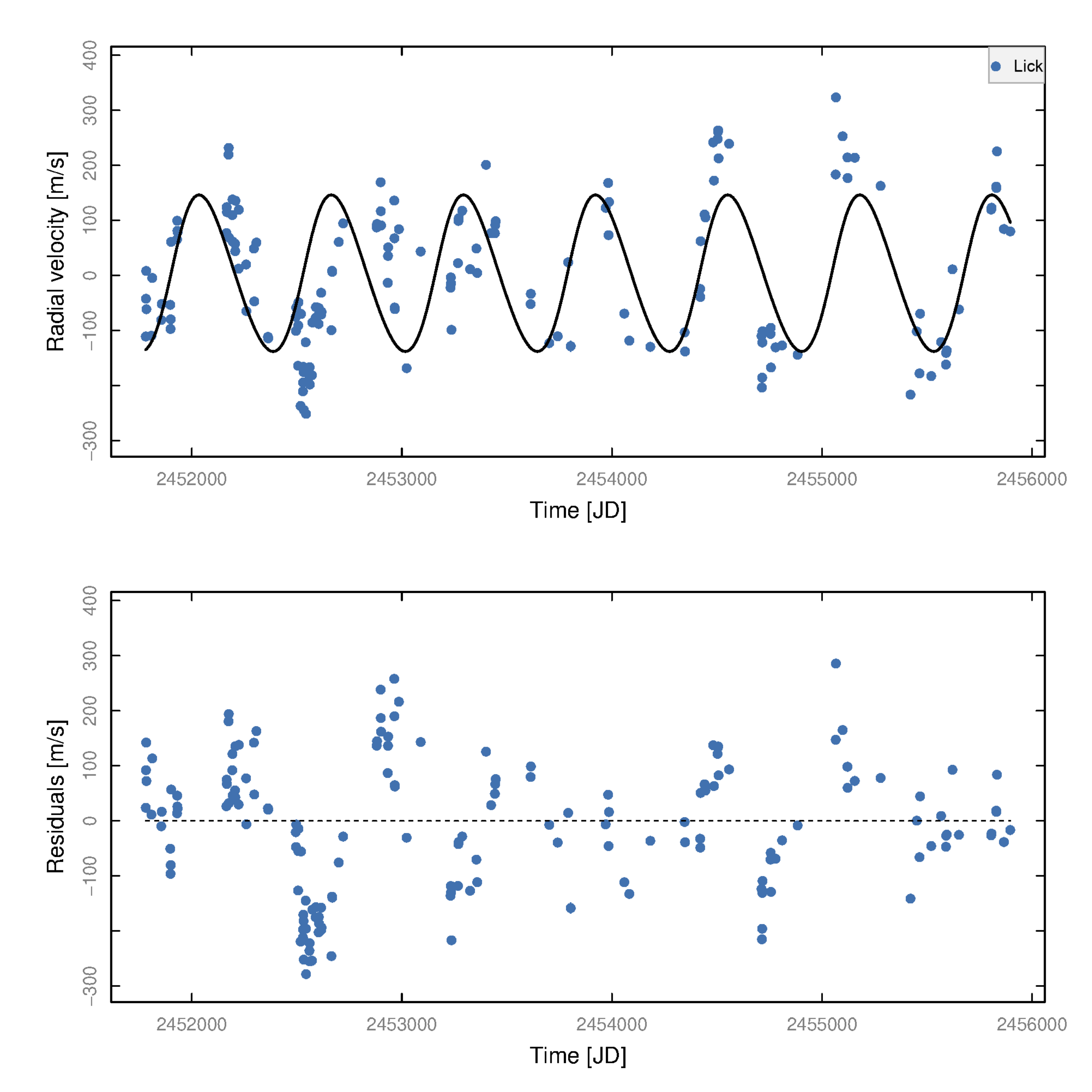}
\caption{\textit{Top left:} RV measurements for Aldebaran from Lick Observatory (165 in total). The orbital solution for a substellar companion with a period of $\sim\!\! \unit[594]{d}$ is represented by the \correction{black} solid line. \textit{Bottom left:} Residual RV variations for the Lick RV measurements of Aldebaran after removing the orbital solution (\correction{black dashed} line) for the Lick data of Table \ref{tab:orbital_sol_comparison}. \textit{Top right:} RV measurements for Aldebaran from Lick Observatory. The orbital solution of H15 for a substellar companion with a period of $\unit[628.96 \pm 0.90]{d}$ is overplotted and is represented by the \correction{black} solid line. \textit{Bottom right:} Residual RV variations for the Lick RV measurements of Aldebaran after removing the orbital solution (\correction{black dashed} line) of H15.}
\label{fig:RV_Lick}
\end{figure*}

\section{The radial-velocity data from Lick Observatory}

According to H15, a companion is the most likely explanation for the $\sim\!\!\unit[629]{d}$ period in the RV measurements. However, it is still unclear whether giant stars show phenomena such as long-term non-radial pulsation patterns; hence, the planet hypothesis should be treated with caution.
If a substellar companion is present, its period should be recovered in independent RV measurements. Therefore, we analysed available RV data from Lick Observatory and searched for RV variations in the range of $ \sim\!\!\unit[629]{d}$. Furthermore, we added those data to the published ones of H15 and conducted the same analyses for the combined data sets as well as for the Lick data separately.

\subsection{The radial-velocity survey at Lick Observatory}

We monitored Aldebaran between 2000 and 2011 using the Hamilton Spectrograph at the 60cm CAT Telescope at Lick Observatory as part of our RV survey of evolved stars \citetext{see, e.g., \citealt{Frink_2001}}. The survey started in 1999 and comprises a sample of 373 G and K giants. Almost all stars are of luminosity class III and brighter than $V=\unit[6]{mag}$. For the survey, the $\unit[0.6]{m}$ Coudé Auxiliary Telescope (CAT) and the Hamilton \'Echelle Spectrograph were used, together with an iodine cell providing a stable reference. In  these data, several extrasolar planets were already discovered \citetext{e.g., \citealt{Frink_2002}, \citealt{Reffert_2006}}. In addition, these data provided valuable insights into the properties of (giant) extrasolar planets around giant stars in a statistical sense \citetext{e.g., \citealt{Reffert_2015}}.

\subsection{The radial-velocity data of Aldebaran}

A total of 165 RV measurements of $\alpha $ Tau were taken at Lick Observatory between $\mathrm{JD}=2451781$ (August 24, 2000) and $\mathrm{JD}=2455896$ (November 30, 2011). The RVs varied between $\unit[-249.1]{ms^{-1}}$ and $\unit[325.5]{ms^{-1}}$ (mean RV subtracted), with a median uncertainty for the measurements of $\sim\!\! \unit[3.8]{ms^{-1}}$. The observed dispersion around the mean of all data points was $\unit[129.3]{ms^{-1}}$. The first 69 measurements in particular are of great importance due to the small number of observations in the data of H15 between early 2000 and late 2002. Unfortunately, even with the complementary Lick data, there is still a large gap of more than six years between August 1993 and January 2000 in which no RV measurements are available.

The aim of combining the data sets of H15 with the Lick data is to obtain a long time baseline with as many measurements as possible in order to extract the long-period RV variation with confidence. The seven data sets of H15 contain 373 RV measurements in total, taken in the time ranges 1980--1993 and 2000--2013. By adding the Lick data we obtain a total number of 538 RV measurements. 

\begin{table*}
\caption{Comparison of the parameters of our orbital solution for the Lick data, and for the combined data, respectively, and those of H15 for the substellar companion to Aldebaran. The corresponding reduced $\chi^2$ and the rms value of the scatter around each fit are listed as well. \correction{The listed offsets for the CFHT, DAO, TLS, BOAO and McDonald data sets are the offsets relative to the offset-corrected published RV data of H15 while the Lick offset refers to the non-offset-corrected RV values in Table~\ref{tab:RV_data_Lick1}.}}
\label{tab:orbital_sol_comparison}
\centering
\begin{tabular}{l r r r}
\hline\hline \vspace{-0.3cm} \\
Parameter & 
\multicolumn{1}{r}{Lick} & 
\multicolumn{1}{r}{Lick + H15} &  \multicolumn{1}{r}{H15} \\
\hline \vspace{-0.3cm} \\
$\unit[P]{[days]}$ & $593.80 \pm 2.75$ & $607.22 \pm 1.17$ & $628.96 \pm 0.90$\\ 
$\unit[T_{\text{Periastron}}]{[JD]}$ & $2451497.4 \pm 15.7$ & $2451491.7 \pm 10.4$ & 
$2451297.0 \pm 50.0$ \\
$\unit[K_1]{[ms^{-1}]}$ & $150.6$ & $157.0$ & $142.1 \pm 7.2$ \\
$e$ & $0.39 \pm 0.06$ & $0.33 \pm 0.04$ & $0.10 \pm 0.05$\\
$\omega$ [deg] & $303.5$ \correction{$\pm  10.5$} & $336.7$ \correction{$\pm 6.5$} & $287 \pm 29$\\
$\chi^2_{\text{red}}$ & 501.74 & 519.55 & 495.98\\
$\unit[\mathrm{rms}]{[ms^{-1}]}$ & 83.26 & 117.37 & 102.13\\
\hline \vspace{-0.3cm} \\
\correction{Offsets $[\mathrm{ms^{-1}}]$} & & & \\
\hline \vspace{-0.3cm} \\
\correction{Lick} & \correction{$2.2\pm 6.7$} & \correction{$4.5 \pm 6.7$} & \correction{\ldots}\\
\correction{CFHT} & \correction{\ldots} & \correction{$9.0 \pm 34.2$} & \correction{\ldots}\\
\correction{DAO} & \correction{\ldots} & \correction{$-10.5 \pm 72.1$} & \correction{\ldots}\\
\correction{TLS} & \correction{\ldots} & \correction{$-35.5 \pm 9.8$} & \correction{\ldots}\\
\correction{BOAO} & \correction{\ldots} & \correction{$-4.3 \pm 18.2$} & \correction{\ldots}\\
\correction{McD-2.1m} & \correction{\ldots} & \correction{$-82.4 \pm 31.3$} & \correction{\ldots}  \\
\correction{McD-CS11} & \correction{\ldots} & \correction{$-21.4 \pm 43.7$} & \correction{\ldots}\\
\correction{McD-Tull} & \correction{\ldots} & \correction{$2.9 \pm 10.7$} & \correction{\ldots}\\
\hline
\end{tabular}
\end{table*}

\section{Analysis of the radial-velocity data}

\subsection{Lomb-Scargle analysis} \label{sec:LS_analysis}

Since our RV data are unevenly sampled, we performed a statistical analysis corresponding to the least-squares spectral analysis of \cite{Lomb_1976} and \cite{Scargle_1982}. Lomb-Scargle (L-S) periodograms showing all trial periods with their statistical power for the Lick data, the seven H15 data sets, and for the combined data sets (Lick + H15), respectively, are shown in Fig.~\ref{fig:Losca_Lick_Hatzes_Comb}.

The most significant peak in the Lick data has a period of $\unit[587.8]{d}$, and a false-alarm probability (FAP) of $1.6 \cdot 10^{-17}$. The period therefore deviates considerably from the most significant period of $\unit[629.1]{d}$ in the middle panel that represents the H15 data. Further periods around $\unit[766]{d}$, $\unit[14]{d}$, $\unit[998]{d}$, and $\unit[263]{d}$ are also apparent in the top panel of Fig.~\ref{fig:Losca_Lick_Hatzes_Comb}. The left part of Table \ref{tab:List_of_Periods} lists all significant periods in the Lick data which have a FAP below $\unit[0.1]{\%}$.

By reanalysing the H15 data we find additional periods which could not be found in the Lick data. Apart from the most prominent period of $\unit[629.1]{d}$, which coincides with the period of the orbital solution for a single substellar companion presented in H15, further RV variations with periods around $\unit[687]{d}$, $\unit[231]{d}$, $\unit[304]{d}$, and $\unit[873]{d}$ occur in the H15 data which are not present in the Lick data.

By adding the Lick data to the H15 data we find that now the most significant peak represents a period of $\unit[625.02]{d}$. The L-S periodogram for the combined data is shown in the bottom panel of Fig.~\ref{fig:Losca_Lick_Hatzes_Comb}, while the exact values of the  periods and their corresponding FAPs are listed in the right part of Table~\ref{tab:List_of_Periods}. The $\unit[766]{d}$ period from the Lick data has slightly increased and is now located at $\unit[775]{d}$. Additional peaks are
visible, for example at $\unit[687]{d}$ and $\unit[873]{d}$, which are lacking in the Lick data and can solely be found in the H15 data. We again have a long-period RV variation near $\unit[1030]{d}$.

\subsection{Orbital solution for the Lick data} \label{sec:Orb_sol_Lick}

Deriving an orbital solution for a single companion for the Lick data yields a period similar to the most significant one from the L-S periodogram in the top panel of Fig. \ref{fig:Losca_Lick_Hatzes_Comb}. However, since the classical L-S periodogram does not take into account eccentric orbits, the values for the periods do not coincide precisely. The fitted solution, which is plotted in the top-left panel of Fig. \ref{fig:RV_Lick}, has a period of $P=\unit[593.80 \pm 2.75]{d}$ and deviates more than $\unit[12]{\sigma}$ from the value of H15 of $P_{\text{H15}}=\unit[628.96 \pm 0.90]{d}$. Taking a stellar jitter of $\sim\!\!\unit[90]{ms^{-1}}$ into account, the reduced $\chi^2$ scales down to almost unity (without jitter, $\chi^2_{\text{red}} = 501.74$). With an eccentricity of $e = 0.39 \pm 0.06$, the orbit of the potential planet is much more eccentric than in the orbital solution of H15 ($e_{\text{H15}} = 0.10 \pm 0.05$) and deviates almost $\unit[4]{\sigma}$ from this value.

In Table \ref{tab:orbital_sol_comparison}, we compare the orbital parameters of the Lick solution (second column) with those of the solution of  H15 (last column). 

Removing the orbital solution reveals a large scatter of the residual RVs around the fit over the whole time range, which can be seen in the bottom-left panel of Fig. \ref{fig:RV_Lick}. The maximum absolute residual RV with a value of $\unit[213.9]{ms^{-1}}$ is reached at $\mathrm{JD}=2451931$ (January 21, 2001). The rms value of the residual RVs is about $\unit[83]{ms^{-1}}$ and therefore of the same order as the expected stellar jitter which is around $\unit[91]{ms^{-1}}$, as estimated from its $B-V$ colour index \citetext{see \citealt{Trifonov_2014}}.

The top-right panel of Figure~\ref{fig:RV_Lick} shows the Lick RV data with the orbital solution of H15 overplotted. The orbital solution of H15 has a reduced $\chi^2_{\text{H15}}$ of $1052.99$ with respect to the Lick data, while our solution has $\chi_{\text{red}}^2=501.74$ (in both cases without accounting for jitter). 

H15 pointed out that there are time spans where their orbital solution does not fit the data points very well due to an apparent phase shift. This can be seen in Fig. 2 of H15, for example between $\mathrm{JD}=2452600$ ($\approx$ November 2002) and $\mathrm{JD}=2453100$ ($\approx$ April 2004).
Such an apparent phase shift around late 2003 and early 2004 between RV measurements and orbital solution is also observable for the Lick data (for our Keplerian fit as well as for the overplotted H15 fit).

This indicates that the signal during that time period is not consistent with a long-lived, coherent one. We can therefore conclude that the Lick data do not further support the hypothesis of a single substellar companion, as they cannot be described very well by a single-planet orbital solution and in particular not by the solution of H15.

\subsection{Orbital solution for the combined data} \label{sec:Orb_sol_comb}

\begin{figure*}
\centering
\includegraphics[width=18cm, trim=1cm 0.5cm 0.5cm 0.9cm,clip]{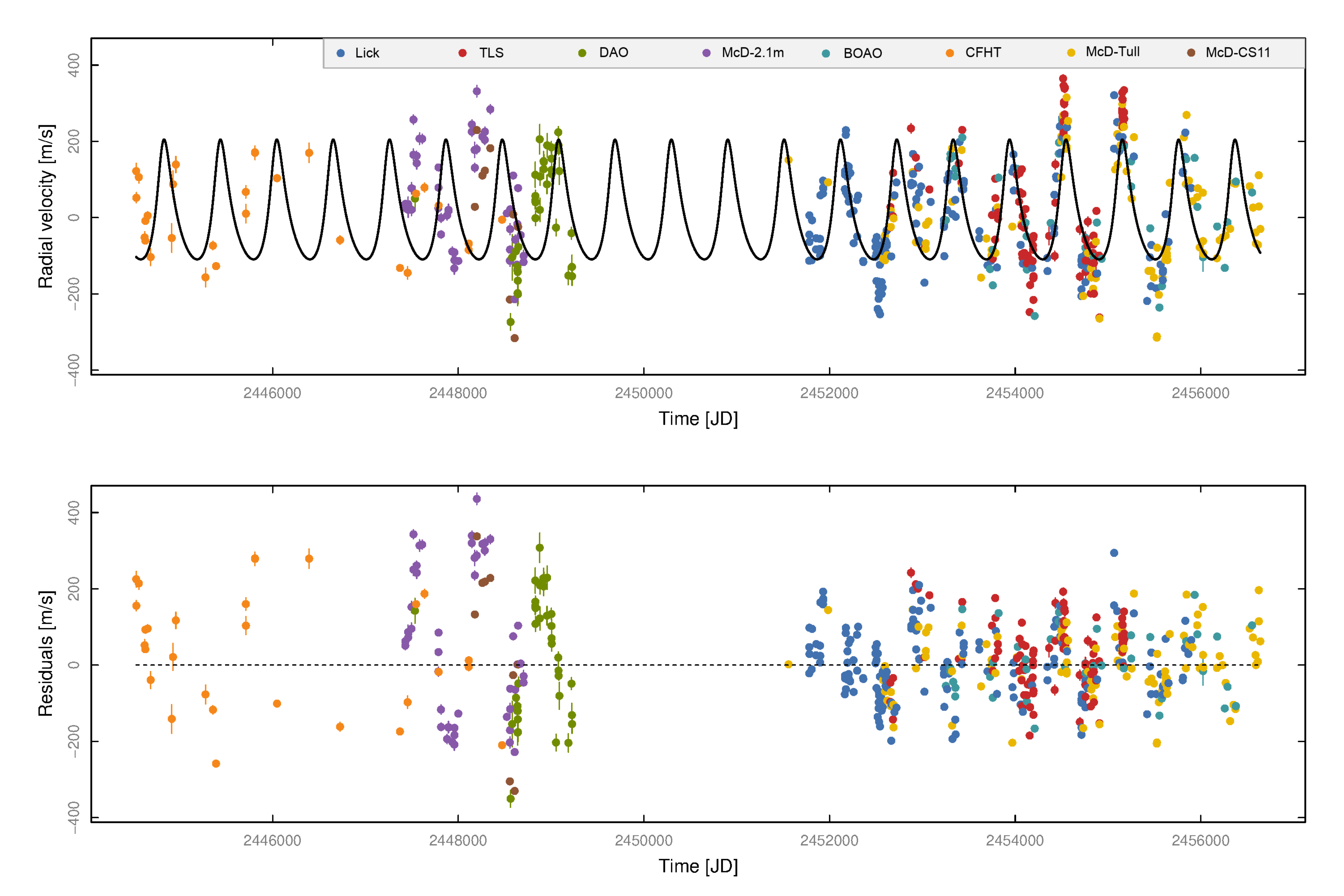}
\caption{RV measurements for Aldebaran from Lick Observatory and all the data sets of H15. The orbital solution for a substellar companion with a period of $\unit[607.22]{d}$ is represented by the \correction{black} solid line \correction{in the top panel; the lower panel shows the residuals with respect to this one-planet fit.}}
\label{fig:RV_all}
\end{figure*}

Figure \ref{fig:RV_all} shows the RVs of the Lick data together with the seven H15 data sets as a function of time. There are still no data available between late 1993 and early 2000. The dispersion around the mean of all data points is $\unit[142.74]{ms^{-1}}$, while the median measurement error is $\unit[4.2]{ms^{-1}}$. The best fitting orbital solution has period $P=\unit[607.22 \pm 1.17]{d}$, eccentricity $e=0.33 \pm 0.04$, RV amplitude $K=\unit[156.98]{ms^{-1}}$, and a reduced $\chi^2$ of $519.55$; a stellar jitter of about 160~m/s would be required for a reduced  $\chi^2_{\text{red}}$ of the order of 1, not too far from the expected value. The value of H15 of $628.96 \pm 0.90$ days deviates more than $\unit[15]{\sigma}$ from the  value determined for the combined data. The discrepancy of the eccentricities (now $e=0.33$, and thus much higher than the H15 value of $e_{\text{H15}}=0.10$) is also not negligible. 

Removing the orbital solution again reveals high residual RVs with a maximum absolute value of $\unit[435.3]{ms^{-1}}$ at $\mathrm{JD}=2448201$ (November 5, 1990), as shown in the bottom panel of Fig.~\ref{fig:RV_all}. The rms value for the residual RVs is $\unit[117]{ms^{-1}}$.

\begin{figure}
\resizebox{\hsize}{!}{\includegraphics{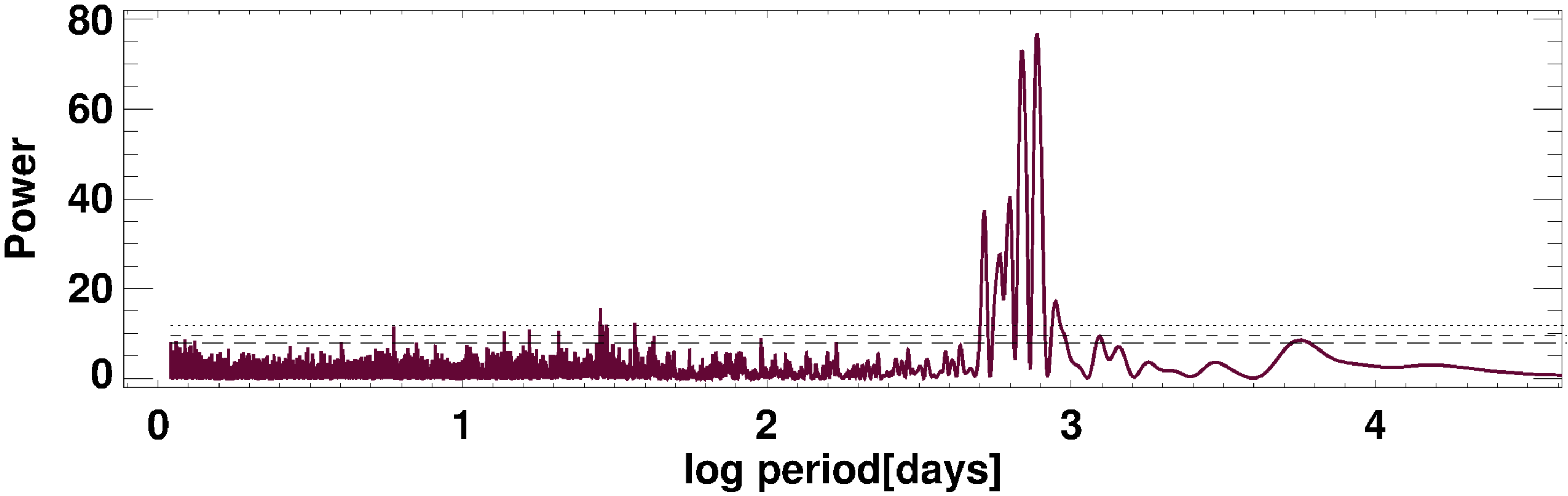}}\caption{Lomb-Scargle periodogram for the Lick data combined with the data of H15 after removing the $\unit[607.22]{d}$ period of the orbital solution. The alias period $\unit[230.66]{d}$ has disappeared as well. The significance thresholds represented by the vertical dashed lines correspond to FAPs of $\unit[0.1]{\%}$, $\unit[1]{\%}$, and $\unit[5]{\%}$, respectively.}
\label{fig:Losca_comb_trendremoved}
\end{figure}

Checking the RV data for alias frequencies is crucial as they often interfere with the originally observed frequencies. For further explanation, see for example \cite{Dawson_2010}.
We find that the $\sim\!\!\unit[620]{d}$ period is not caused by a linear combination of other frequencies, but that it is responsible for the 230~day peak in the L-S periodogram ($f_{230d} \approx f_{1yr} + f_{620d}$). Removing the orbital solution therefore causes the $\unit[230.66]{d}$ period to disappear (see Fig.~\ref{fig:Losca_comb_trendremoved}). Very significant peaks at $\unit[775]{d}$ and $\unit[687]{d}$ are still present. A period around $\unit[873]{d}$ is also still observable in Fig.~\ref{fig:Losca_comb_trendremoved} above the significance level, although its power decreased relative to the highest peaks. 
It might be the corresponding alias to the 230d period, since
$f_{873d} \approx f_{1yr} - f_{620d}$. The same can be observed for the Lick data (left part of Tab. \ref{tab:List_of_Periods}), where the $\unit[225]{d}$ and $\unit[988]{d}$ periods could be $\unit[1]{yr}$ aliases of the dominant $\unit[588]{d}$ period ($f_{225d} \approx f_{1yr} + f_{588d}$ and $f_{998d} \approx  f_{1yr} - f_{588d}$).

\subsection{Growth of the Lomb-Scargle power}

\begin{figure}
\resizebox{\hsize}{!}{\includegraphics{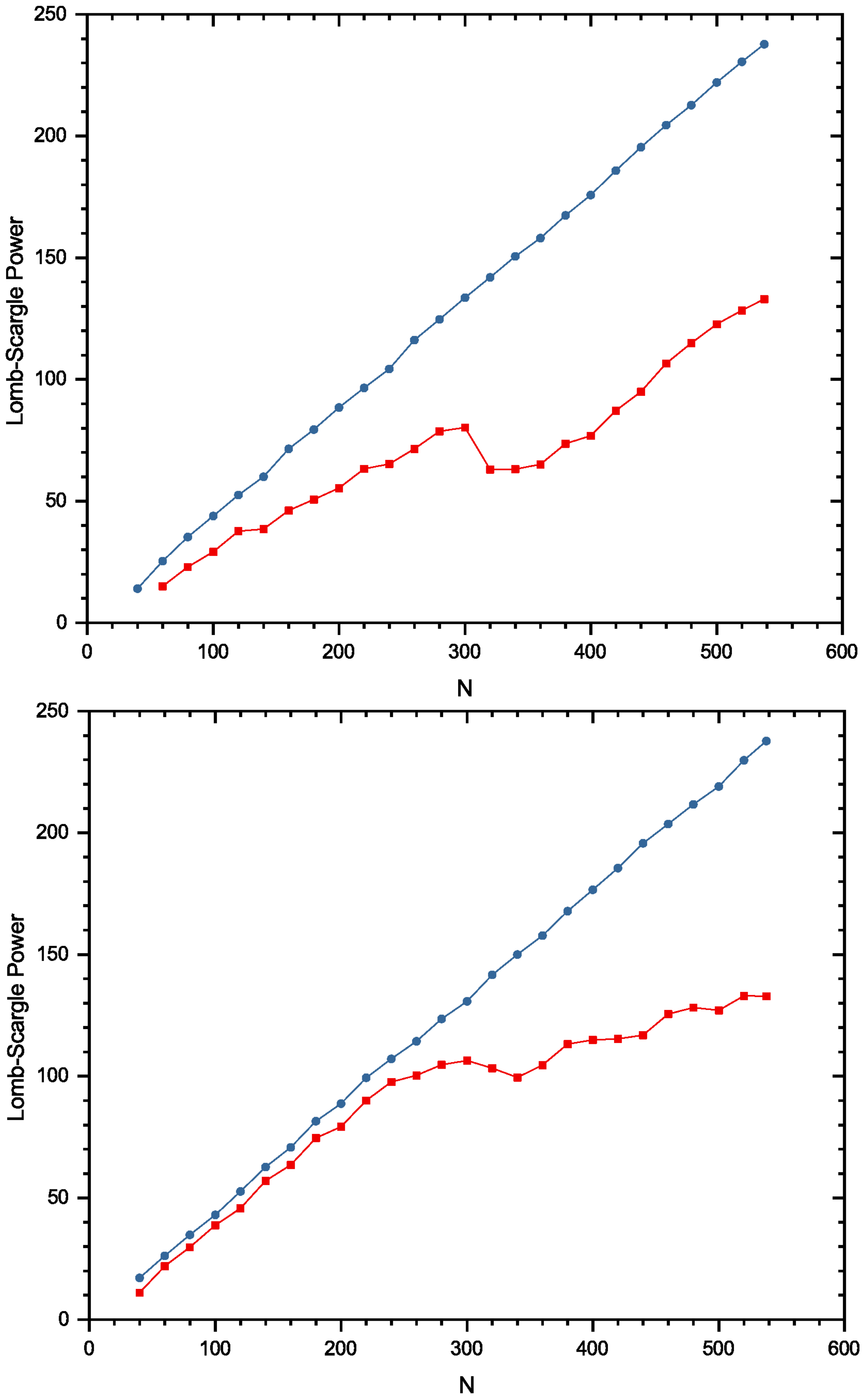}}
\caption{Lomb-Scargle power of the $\unit[625]{d}$ period as a function of the number of data points ($N$). The red squared data points represent the real data while the blue circular data points represent simulated data. \textit{Top:} Growth of the Lomb-Scargle power from 1980--2013. Data points were added in steps of 20 in chronological order. A decrease in statistical power is observable for the real data between $N=300$ and $N=320$. \textit{Bottom:} Growth of the Lomb-Scargle power from 2013--1980. Data points were added in steps of 20 in reversed chronological order. Before $N=240$ (which represent the newest RV data) the $\sim\!\! \unit[620]{d}$ period is clearly visible as the growth in power shows the same linear slope as the simulated data. The older RV data (taken before 2006) however cause a decrease in the slope and a drop in power at $N=340$.}
\label{fig:Growth_Power_Losca}
\end{figure}

An exoplanet orbiting a star should cause a strictly periodic RV signal whose statistical L-S power increases linearly with the number of measurements. For simulated data sampled in the same manner as the actual \correction{538} RV measurements, this is shown by the blue circles in both panels of Fig. \ref{fig:Growth_Power_Losca}. The simulated data have the same measurement errors as the real data, and a stellar jitter of $\unit[91]{ms^{-1}}$ has been added.

The actual measured data from H15 and the Lick survey, added in chronological order and represented by the red squares in the top panel of Fig.~\ref{fig:Growth_Power_Losca}, have a shallower slope, and at some point show a non-linear behaviour; the L-S power temporarily decreases between the 300th and the 320th data point, corresponding to the time range between November 2006 and February 2007. In Fig.~5 in H15, such a decrease in power is also observable in the same time span (we note that H15 erroneously assign it to the time range between 2002 and early 2004, where they observed an apparent phase shift between the RV measurements and their orbital solution). The systematic decrease in L-S power during this time span weakens the hypothesis of a substellar companion.

Adding the data in reverse chronological order (see bottom panel of Fig.~\ref{fig:Growth_Power_Losca}) reveals that the newest data ($\sim\!\! 2006 - 2013$), which represent in this panel $N =1$ to $N= 240$, show a coherent RV variation in the range of $\unit[620]{d}$; the L-S power linearly increases with the number of measurements. After $N=240$ the slope decreases significantly, which is a hint that the $\unit[620]{d}$ signal is not as clearly present in the data taken before 2006 as it should be if it is caused by a substellar companion. 

\begin{figure}
\resizebox{\hsize}{!}{\includegraphics{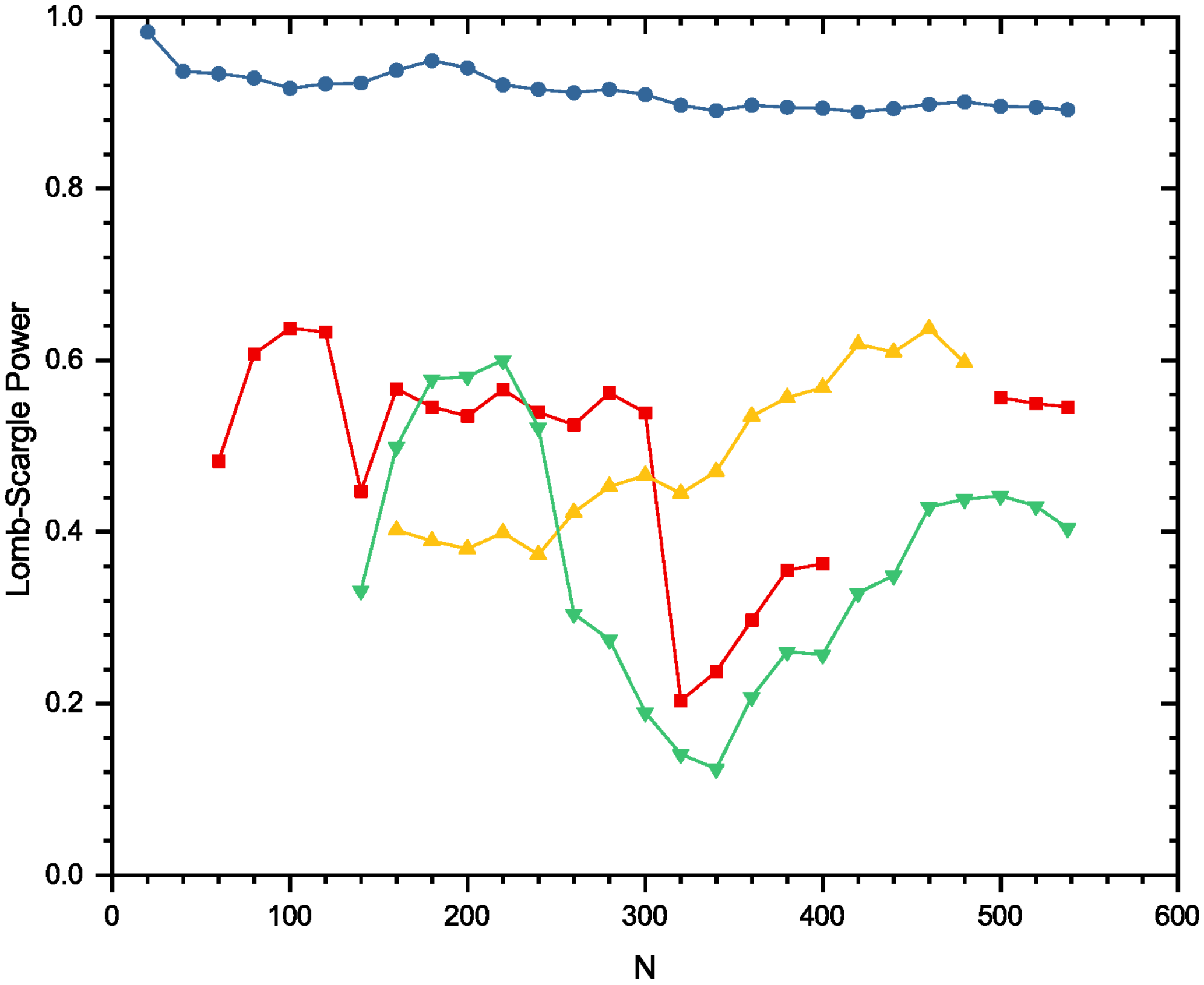}}
\caption{Growth of the L-S power in the GLS as a function of the number of data points ($N$). Data points were added in steps of 20 in chronological order. The red squared data points represent the $\sim\!\!\unit[620]{d}$ period, while the green triangles represent the $\sim\!\!\unit[750]{d}$ period and the yellow triangles the $\sim\!\!\unit[550]{d}$ period. The blue circular data points represent simulated data. See the text for more details.}
\label{fig:Growth_Power_GLS_each_period}
\end{figure}

Since the generalized Lomb-Scargle periodogram (GLS) of \cite{Zechmeister_2009} has a different normalization (it is normalized to unity) and can show a slightly different behaviour we compare the growth of the power in the classical L-S periodogram (from the top panel in Fig. \ref{fig:Growth_Power_Losca}) with the growth of power in the GLS. While for the combined data the L-S power remains almost unity for the simulated data, it shows multiple strong drops for the real data, which can be seen in Fig. \ref{fig:Growth_Power_GLS_each_period}. The most significant period in the GLS even changes while adding further RV measurements, which is indicated by the different colours. After the first 60 data points, the period in the range of $\sim\!\!\unit[620]{d}$ arises and is most significant. After 140--160 measurements, two further periods around $\unit[550]{d}$ and $\unit[760]{d}$ become dominant. The latter one replaces the $\sim\!\!\unit[620]{d}$ peak as most significant period from $N=180$ to $N=240$, while the $\sim\!\!\unit[550]{d}$ RV signal dominates the periodogram between $N=320$ and $N=500$. Although the $\sim\!\!\unit[620]{d}$ period completely vanishes between $N=400$ and $N=500$ (most likely due to its proximity to the $\sim\!\!\unit[550]{d}$ period), it again becomes the most significant period after 500 measurements.

\subsection{Orbital solution with two planetary companions}

\begin{figure*}
\centering
\includegraphics[width=18cm, trim=1cm 0.5cm 0.5cm 0.9cm,clip]{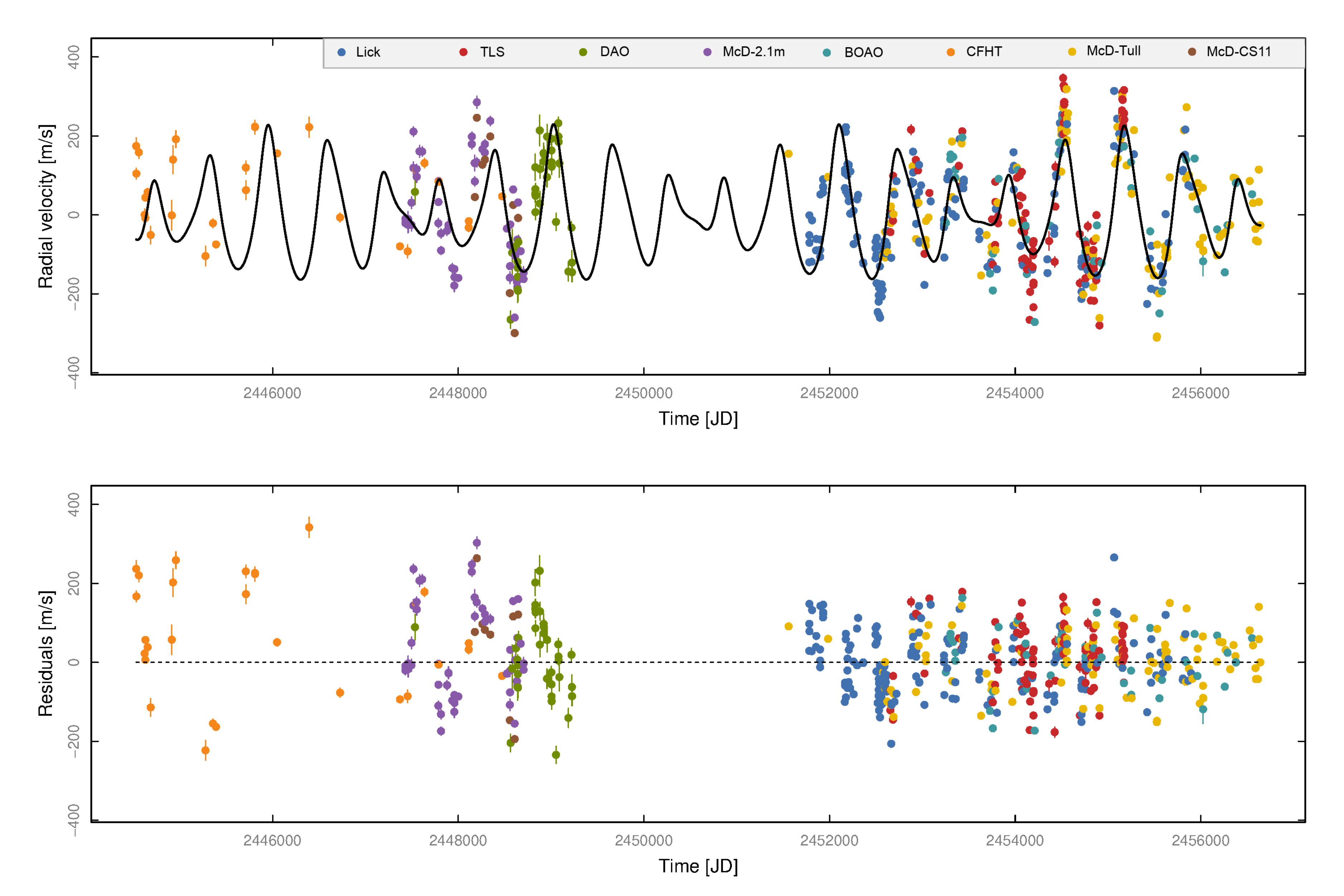}
\caption{RV measurements for Aldebaran from Lick Observatory and all data sets of H15. The orbital solution for two substellar companions with periods of $\unit[614.10]{d}$ and $\unit[772.83]{d}$ is represented by the \correction{black} solid line in the top panel; the lower panel shows the residuals with respect to this two\correction{-}planet fit.}
\label{fig:RV_all_2}
\end{figure*}

As we still have two very significant periods ($\unit[775]{d}$ and $\unit[687]{d}$) left in Fig. \ref{fig:Losca_comb_trendremoved}, it is worth also investigating solutions with multiple planets. The best Keplerian two-planet solution we find has orbital periods of $\unit[614.10]{d}$ and $\unit[772.83]{d}$, where the period of the potential outer planet corresponds to the most significant period of Fig. \ref{fig:Losca_comb_trendremoved}. This two-planet solution for all eight data sets is shown in the upper panel of Fig. \ref{fig:RV_all_2}. Especially for the old RV data taken between 1980 and 1993, the scatter around the two-planet fit is still very large, as can be seen in the lower panel of Fig. \ref{fig:RV_all_2}. The orbital parameters of both potential planets as well as the reduced $\chi^2$ and the rms scatter of the residual RVs around the two-planet fit are listed in Table \ref{tab:orbital_sol_2_planets}.

However, as both planets would have several Jovian masses and quite similar semi-major axes it is unlikely that this corresponds to a stable configuration. A dynamical fit to the data was not possible as the fit did not converge, or the solution was not stable over the timescale of the observations. In order to find possible dynamically stable solutions we create \correction{11} million samples of orbital parameters for both planets\correction{, 10 million samples} based on Gaussian distributions
around the 3$\,\sigma$ parameter space of each of our Keplerian best-fit parameters, namely period, mean anomaly, eccentricity, longitude of periastron and RV semi-amplitude. \correction{An additional 1 million samples were uniformly sampled around the 10$\,\sigma$ parameter space of each of our Keplerian best-fit parameters in order to find possible dynamically stable islands which in case of significant dynamical interaction could lie far away from the best Keplerian fit parameters.} The errors of our Keplerian best-fit parameters for the hypothetical two\correction{-}planet solution that were used for the Monte-Carlo (MC) sampling are determined by bootstrapping and are slightly larger than the errors based on the covariance matrix shown in Table~\ref{tab:orbital_sol_2_planets}. 
\correction{The bootstrap uncertainties for all solutions presented in this paper are given in
Table~\ref{tab:orbital_sol_comparison_bootstrap} in the Appendix.} 

We use co-planar configurations with $i=90^\circ$ and keep the mass of the host star fixed at $\unit[0.91]{M_\odot}$. We test all \correction{11} million samples for long-term dynamical stability using the Wisdom-Holman integrator \citep[also known as MVS;][]{Wisdom1991}, which was modified to work with Jacobi input elements. We choose 1~Myr as the maximum integration time as well as an integration time step of 1~day. The fit is regarded unstable if at any time the separation between the star and one of the companions exceeds $\unit[5]{AU}$ (planet ejection) or falls below $\unit[0.001]{AU}$ (star-planet collision, since this limit is well below the stellar radius). 

We find only a very small number of dynamically stable solutions around our best Keplerian fit. Out of the 10 million \correction{Gaussian sampled} combinations of orbital parameters investigated only $2016$ result in stable configurations according to our definition, which corresponds to a fraction of only $\unit[0.02]{\%}$. \correction{From the 1 million uniformly sampled combinations of orbital parameters around the 10$\,\sigma$ parameter space of the best Keplerian fit we find 455 stable configurations ($\unit[0.05]{\%}$). These} number\correction{s} would further decrease if we allow inclination angles less than $90^\circ$.  

Despite the small fraction of stable configurations, these represent an interesting possibility for a stable two-planet system. To see if these configurations could explain the observed RVs we reconstruct the radial stellar motion induced by these synthesised stable configurations over the past $\sim\!\!\unit[35]{yr}$ and we test it against the data. For this purpose we adopted a dynamical model \citep{Tan_2013,Trifonov_2014} fed with fixed initial orbital parameters from the stable samples and adjustable RV offsets for each dataset. By studying the quality and the overall dynamics of these models, we find the following:\\
(i) Twelve out of the 2016 \correction{Gaussian sampled} stable solutions are in a 4:3 MMR, but the quality of these models is poor as they show rather large residuals. The rms of the residuals of the best 4:3 MMR model is $\unit[127.72]{ms^{-1}}$, which is larger than the rms of the simpler one-planet model for the combined data, which has $\unit[117.37]{ms^{-1}}$.\\
(ii) The rest of the \correction{Gaussian sampled} stable solutions were found in 1:1 MMR configurations, which induce an RV curve practically indistinguishable from a one-planet model \citep[e.g. see][]{Laughlin_2002}. Therefore, the quality of these considerably more complex 1:1 MMR two-planet models is comparable, or in most cases even worse than the one-planet model, which once again speaks against the two-planet hypothesis.\\
\correction{(iii) Investigating the dynamical models of the 455 stable solutions, which were uniformly sampled around the 10$\,\sigma$ parameter space of the best Keplerian fit, reveals that these fits are not consistent with the observed RV data. For the vast majority of stable solutions, the eccentricity of the potential second planet ($P=772.83$ d) is higher than $e=0.4$, resulting in RV fits that describe the observed RV variations poorly and show clear systematics in the residuals. The rms of the best-fitting dynamical model out of the 455 stable solutions is $\unit[116.9]{ms^{-1}}$, which is no significant improvement compared to the simpler one-planet model.}

Further, almost all of the stable solutions have crossing orbits; it is rather unclear whether and how such a system could form. Given the tiny fraction of stable solutions in our MC sample\correction{s}, the presence of crossing orbits in almost all stable solutions, the presence of systematics in the residuals of the RV data, and the poor quality of these stable configurations, we conclude that the two-planet model is not a viable solution to explain the RV data of Aldebaran.

\begin{table}
\caption{Planetary parameters of the orbital solution for two substellar companions to Aldebaran for the Lick + H15 data sets. The reduced $\chi^2$ and the rms value of the scatter around the Keplerian fit are listed as well.}
\label{tab:orbital_sol_2_planets}
\centering
\begin{tabular}{l r r}
\hline\hline \vspace{-0.3cm} \\
Parameter & Planet 1 & Planet 2\\
\hline \vspace{-0.3cm} \\
$\unit[P]{[days]}$ & $ 614.10 \pm 1.18 $ & $ 772.83 \pm 4.34 $ \\ 
$\unit[T_{\text{Periastron}}]{[JD]}$ & $ 2452688.1 \pm 15.8 $ & $ 2452930.7 \pm 98.6 $\\
$\unit[K_1]{[ms^{-1}]}$ & $ 125.42 $ & $ 72.24 $ \\
$e$ & $ 0.24\pm 0.04 $ & $ 0.09 \pm 0.08$ \\
$\omega$ [deg] & $ 336.2$ \correction{$\pm 9.7$} & $  24.4 $ \correction{$\pm 46.4$} \\
$\chi^2_{\text{red}}$ & \multicolumn{2}{c}{358.97}  \\
$\unit[\mathrm{rms}]{[ms^{-1}]}$ & \multicolumn{2}{c}{89.65} \\
\hline \vspace{-0.3cm} \\
\correction{Offsets $[\mathrm{ms^{-1}}]$} & & \\
\hline \vspace{-0.3cm} \\
\correction{Lick} & \multicolumn{2}{c}{\correction{$11.4 \pm 5.9$}}\\
\correction{CFHT} & \multicolumn{2}{c}{\correction{$-43.4 \pm 34.4$}}\\
\correction{DAO} & \multicolumn{2}{c}{\correction{$-19.0 \pm 60.0$}}\\
\correction{TLS} & \multicolumn{2}{c}{\correction{$-17.3 \pm 8.5$}}\\
\correction{BOAO} & \multicolumn{2}{c}{\correction{$9.4 \pm 15.2$}}\\
\correction{McD-2.1m} & \multicolumn{2}{c}{\correction{$-36.3 \pm 26.2$}}\\
\correction{McD-CS11} & \multicolumn{2}{c}{\correction{$-38.1 \pm 36.4$}}\\
\correction{McD-Tull} & \multicolumn{2}{c}{\correction{$-0.7 \pm 8.9$}}\\
\hline
\end{tabular}
\end{table}

\begin{figure*}
\centering
\includegraphics[width=18cm]{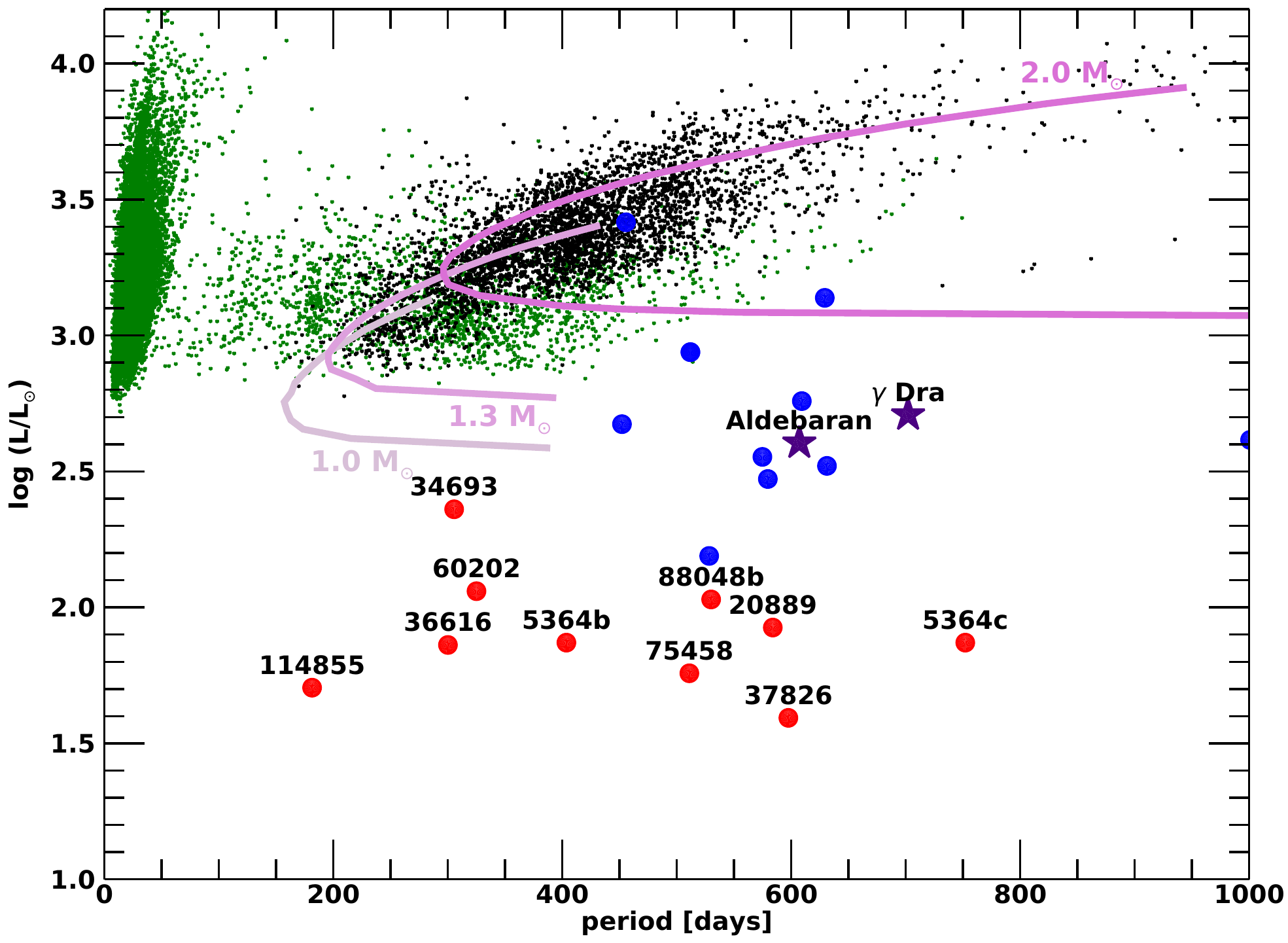}
\caption{\label{plum} Period-luminosity diagram to illustrate the position of Aldebaran and $\gamma$~Dra (star symbols) with respect to published exoplanets from our Lick survey (red circles, labeled with HIP number) and candidate exoplanets (blue circles). The small dots are stars with periods identified by the OGLE survey in the LMC; the subset of black dots are those stars identified as showing long secondary periods in the OGLE photometry. The thick solid lines show model calculations by \citet{Saio_2015} for the periods of oscillatory convective modes for stars of various masses, and for a mixing length parameter of 1.2. Aldebaran and $\gamma$~Dra's periods are rather close to the predicted periods of oscillatory convective modes, especially if one were to extrapolate the models for 1.0 and 1.3~M$_{\odot}$.}
\end{figure*}

\section{Oscillatory convective modes?}

Long secondary periods, which are of the order of 400--1500 days and are found in about
25--30\% of AGB stars \citep{Wood_2004}, have been known for a long time, but not
theoretically understood. \citet{Saio_2015} suggested that non-adiabatic g$^-$ modes present in the deep convective envelopes of luminous giant stars, so-called oscillatory convective modes, could be responsible for the observed periods.
They have presented calculations which show that, depending on the adopted value for the mixing length parameter, oscillatory convective modes would be present in the same region of period-luminosity space where the stars with long secondary periods are found. 

In order to test whether oscillatory convective modes could possibly explain the observed RV variations in Aldebaran, we compare its location in period-luminosity space with those of other stars known to exhibit long secondary periods, similar to the diagram presented by \cite{Saio_2015}, in Fig.~\ref{plum}.
The small dots represent stars with photometric periods determined from the OGLE survey of the Large Magellanic Cloud \citep[LMC;][]{Soszynski_2009}. The stars have been identified in 2MASS and dereddened, so that their extinction corrected colours best match expected intrinsic colours of giant stars as given by \cite{Bessell_1988}. 
The scaling factors for the absorption in the various photometric bands were taken from \citet{Schlegel_1998}. We then used the infrared flux method \citep{Blackwell_1994} to obtain luminosities. 

We only show in Fig.~\ref{plum} those stars for which the 2MASS photometry had errors smaller than 0.05 mag in K band, and for which the absorption could be relatively well determined (good match to expected intrinsic colours after correction for absorption, and formal error of the fitted absorption less than 0.5 mag in V band). These criteria eliminated many of the initial 91\,644 stars in the OGLE sample with 2MASS photometry, leaving 55\,832 stars. Of those, 32\,601  have photometric amplitudes smaller than 0.015 mag and are shown as green points in Fig.~\ref{plum}. 
We choose an upper limit on the photometric amplitude, since the stars in our Lick survey are known not to be photometrically variable based on Hipparcos photometry. 
The subset of black points are those 3833 stars which have been identified by \citet{Soszynski_2009} to fall on sequence~D, that is,\ those stars with long secondary periods. Whether or not the Lick survey stars are photometrically variable with periods similar to these sequence-D stars and small photometric amplitudes (mmag level) is not known, since the cadence of the Hipparcos photometry would not have allowed this kind of photometric periodicity to be detected. 

We overplotted the computed periods for stars of 1.0, 1.3, and 2.0 M$_{\odot}$ and
various luminosities from \citet{Saio_2015} (solid lines), for a mixing length parameter of 1.2. These periods coincide in general with the order of magnitude of the observed periods and stellar luminosities in the OGLE sample. A mixing length parameter value of 1.9 would provide a poorer fit to the sequence-D stars (black), but in turn a better fit to the radial modes of AGB stars on sequences A, B, C, and C$^{\prime}$ (green points). Since the models by \citet{Saio_2015} for the calculation of oscillatory convective modes
do not take into account turbulent pressure and possible overshooting, the precise shape and location of the models could change. In particular, \citet{Saio_2015} note that a better theory of convection could extend the range in which oscillatory convective modes are found. 

Furthermore, we note that the theoretical calculations of the oscillatory convective mode periods have been performed for stellar models with slightly sub-solar metallicity (Z=0.008, corresponding to [Fe/H]$\approx -0.3$~dex). For the comparison with data from the LMC, one should take into account that the average metallicity is somewhat smaller in the LMC than in the solar neighbourhood; \citet{Cole_2000} for instance determined an average metallicity of around $-0.6$~dex for a sample of field giants in the LMC. On the other hand, the median metallicity of stars in the Lick sample is $-0.09$~dex, while the stars orbited by giant planets (and shown in Fig.~\ref{plum}) have preferentially higher metallicities.
Thus, we caution that different underlying metallicities could affect the comparison between data and theory in Fig.~\ref{plum} to some extent.

We also indicate in Fig.~\ref{plum} the locations of Aldebaran and $\gamma$~Dra (star symbols), as well as published exoplanets (red circles) from our Lick survey. Blue circles indicate stars with more or less periodic RV patterns with uncertain interpretation at the moment; the RV variations could either be caused by an orbiting planet or by non-radial pulsations. There is a clear separation in luminosity visible between those categories: published planet discoveries from our survey are typically found around stars with $\log(L/L_{\odot})$ smaller than 2.5 ($L<300\,L_{\odot}$), while the more doubtful cases have $\log(L/L_{\odot})$ larger than 2.5. It would be tempting now to attribute this to the onset of oscillatory convective modes for the higher-luminosity stars, but it could also be the result of the well-known fact that higher-luminosity stars show larger RV jitter. This is very well understood and is due to solar-like (i.e.\ radial) pulsations, with periods much smaller (typically hours do days) than the long secondary periods discussed here. The larger RV jitter makes it harder to distinguish true exoplanets with Keplerian RV curves from less regular RV curves which show periods, amplitude, and/or phase changes, such as Aldebaran. In any case, a much larger number of RV measurements taken over a longer time period is necessary for an unambiguous planet detection around evolved stars of high luminosity.

\citet{Hatzes_2018} already noted that the period of  $\gamma$~Dra is very close to the period predicted
for oscillatory convective modes by \citet{Saio_2015}, and the same is true for Aldebaran, which has
very similar stellar parameters and evolutionary stage to $\gamma$~Dra. Although we presently have no way
to unambigously confirm or reject oscillatory convective modes in Aldebaran, Fig.~\ref{plum}
cautions us to be very careful with exoplanet claims around stars with luminosities larger than
about 300\,L$_{\odot}$.

\section{Discussion and conclusion} \label{sec:Discussion_and_Conclusion}
The number of confirmed detections of extrasolar planets around giant stars is small compared to the total number of confirmed extrasolar planets. Therefore, it is of special interest to have a clean sample as a basis for proper statistics of planets around giant stars. Here we have further investigated the evidence for a Jupiter-mass planet around the K giant Aldebaran, which had been suggested by \cite{Hatzes_2015}.

We conclude that the RV data from Lick Observatory do not support the existence of this planet. By fitting an orbital solution to the Lick data combined with those of \cite{Hatzes_2015} we obtain large residual RVs up to $\unit[435.3]{ms^{-1}}$. Even if we take into account the fact that Aldebaran is an evolved K giant with a stellar jitter in the range of $\unit[100]{ms^{-1}}$, this large systematic scatter around the orbital solution weakens the planet hypothesis.

Plotting the power of the most significant period in the Lomb-Scargle periodogram while continuously adding further measurements in chronological order reveals a strong decrease in statistical power after 300 measurements, contrary to expectations for a strictly periodic signal. Using the generalized L-S periodogram instead yields several instances where the power declines, and further reveals that the claimed planet period is not continuously present; it changes considerably in significance depending on the number of measurements.

A system with two planets formally fits the RV data better than a single planet. However, both planets would have at least several Jupiter masses and similar semi-major axes (around $\unit[1.5]{AU}$), leading to large gravitational interactions. We perform a dynamical analysis and search for stable solutions in the vicinity of the best Keplerian fit, integrating for up to  $\unit[1]{Myr}$. We find that only $\unit[0.02]{\%}$ of \correction{the Gaussian sampled} \correction{and only $\unit[0.05]{\%}$ of the uniformly sampled} configurations are stable over this time range, and many of these formally stable solutions involve crossing orbits. We thus conclude that a two-planet configuration cannot explain the RV observations of Aldebaran.

The reason for the observable RV variations remains unknown, but an intrinsic effect related to the star itself appears more likely than an extrinsic one. In general, long-period variations in K~giants could be caused by non-radial pulsations, spots (rotational modulation of surface features), or substellar companions. Since Aldebaran is photometrically stable at the level of $\unit[10]{mmag}$ \citep{1997ESASP1200.....E}, large stellar spots that could in principle explain the large-amplitude RV variations cannot be present. Radial pulsations in K giants take place at timescales that are much too short to explain the long periods present in the periodograms; based on the scaling relations from \cite{Kjeldsen_1995} the radial pulsation period for Aldebaran should be about 6.6 days. 

An apparent phase shift between the RV data and the orbital solution, for example around 2004, as described in \cite{Hatzes_2015}, is also observable in the Lick data as well as in the combined RV data sets. \cite{Hatzes_2018} recently described a similar phenomenon for the K5 III star $\gamma$~Dra, where RV measurements taken between 2003 and 2011 suggest the presence of a planetary companion. New measurements however taken between 2011 and 2017 show that the RV variations abruptly cease around 2011 and return phase-shifted after three years with the same amplitude and period. The authors argue that this could possibly be related to dipole oscillatory convective modes, as proposed by \cite{Saio_2015}. If this is indeed the case for $\gamma$~Dra, it could also be considered a likely explanation for Aldebaran, as the two stars are remarkably similar (both have spectral type K5 III, $T_{\mathrm{eff}}$ between $3900$ and $\unit[4000]{K}$, $\log(\unit[g]{[\mbox{cm/s}^2]}$) between 1.2 and 1.7 and $\log(L/\mathrm{L}_{\odot})$ between 2.6 and 2.7, and both stars are nearing the tip of the RGB and will start burning helium soon). Since the predicted period-luminosity relations for dipole oscillatory convection modes roughly match the observations of $\gamma$~Dra, the same is true for Aldebaran (although the mass of Aldebaran is slightly smaller than that of $\gamma$~Dra). 

We cannot completely rule out the existence of one or multiple substellar companions around the K giant star Aldebaran. However, it seems that neither a single- nor a two-planet solution is a sufficient explanation for the observed RV variations. One could however imagine that stellar variability is superimposed on the RV signal of one planet, but it would still be difficult to explain observed phenomena such as the apparent phase shift around 2004 between the RV data and the Keplerian fit. Nevertheless, the most likely cause seems to be a type of stellar oscillation in highly evolved K giant stars that we have not yet discovered. Further investigations on the nature of giant stars, especially of the evolved K5 III-type giants, are therefore needed before conclusions can be made on the true cause of the RV variations in giant stars like Aldebaran.

\begin{acknowledgements} We kindly thank Hideyuki Saio for making his computed periods of oscillatory convective modes available to us. We also want to thank Paul Heeren for his technical assistance with some parts of the dynamical analysis. \correction{This work makes use of \textit{Systemic2} \citep{Meschiari_2009}.
K.R.\ and S.R.\ acknowledge support by the DFG Priority Program 'Exploring the Diversity of Extrasolar Planets' (SP~345/20-1 and RE~2694/5-1, respectively).} S.S.\ and S.R.\ \correction{further }acknowledge support by the DFG Research Unit FOR2544 'Blue Planets around Red Stars' (RE~2694/4-1).
\end{acknowledgements}

\bibpunct{(}{)}{;}{a}{}{,} 
\bibliographystyle{aa} 
\bibliography{Reichert_et_al_2019.bib}

\begin{appendix}
\section{Comments on \cite{Farr_2018}}
\label{appa}

\cite{Farr_2018} fitted a planet to all the available RV data sets of H15 plus
their own high-cadence SONG data, while simultaneously applying a Gaussian Process-based Continuous Auto-Regressive Moving Average (CARMA) method to model the acoustic oscillations which happen on much shorter timescales. The planet period which they fitted is 629 days, while the stellar oscillations have their maximum frequency at 2.24\,$\mu$Hz, corresponding to a period of 5.1~days \citep{Farr_2018}.

\finalcorrection{It is of course possible to model any signal with the CARMA formalism. However, the stochastic component must then contain power at the periods of the residuals to the Keplerian component. This is apparently why \citet{Farr_2018} achieved a good fit to the data, but is inconsistent with their interpretation of the stochastic component with short-period oscillations as we will show in the following.} 

\begin{figure*}
\centering
\includegraphics[width=18cm]{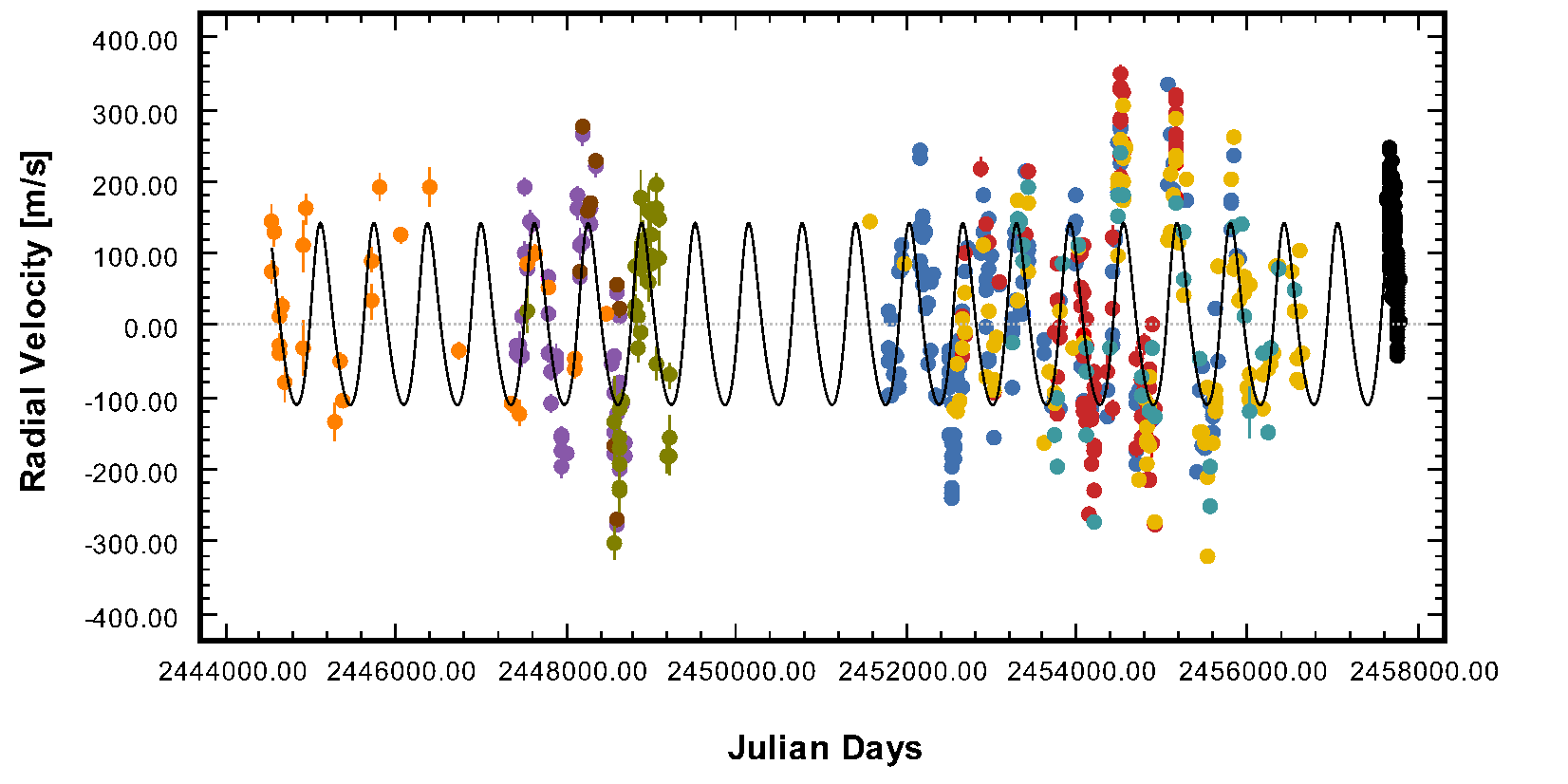}
\includegraphics[width=18cm]{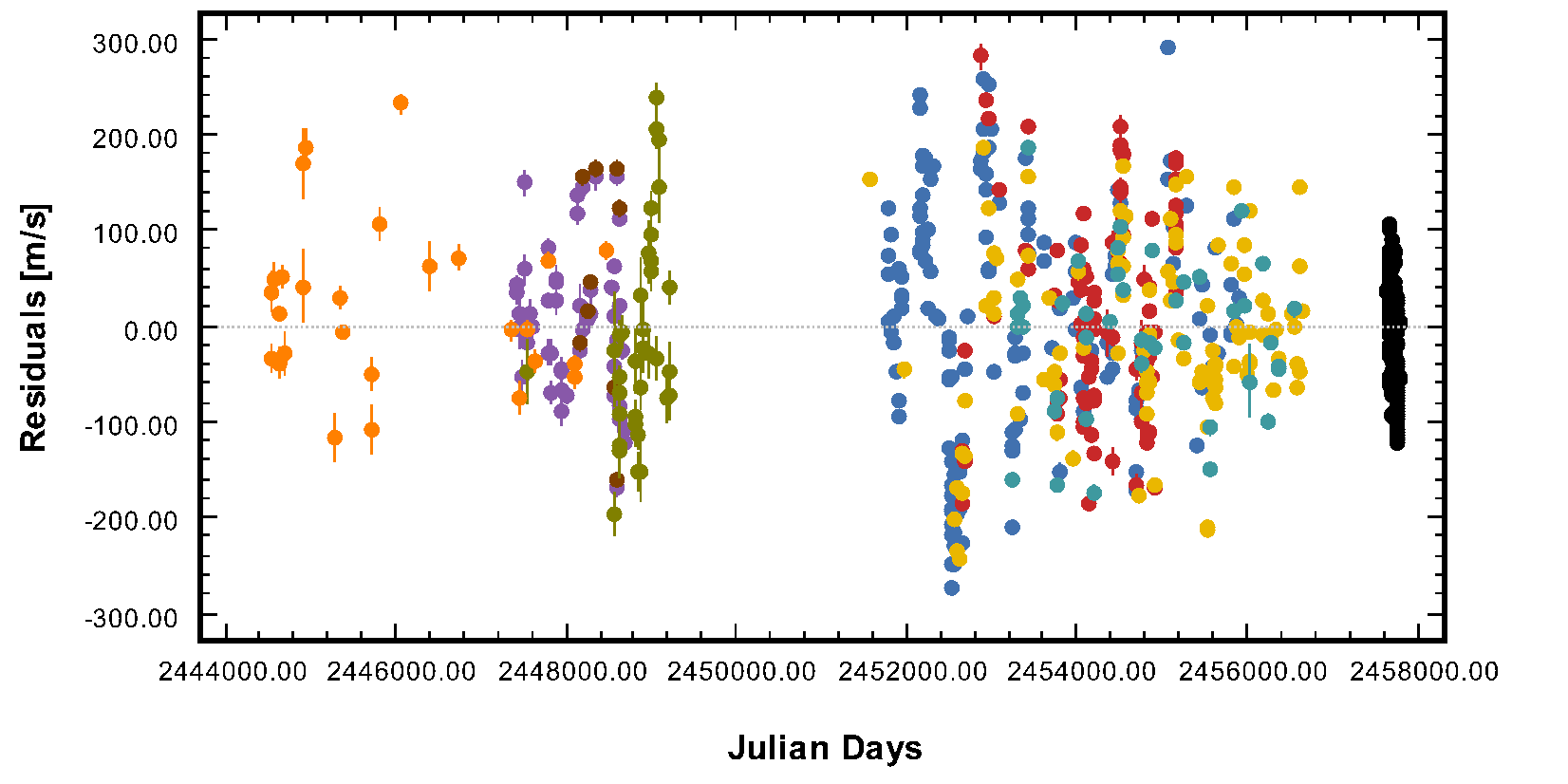}
\caption{\label{farr_fit} All available RV data
sets for Aldebaran, i.e.\ all the H15 data sets plus the ones from Lick (in blue) and Song (in black), with the planet fit from \cite{Farr_2018} overplotted in the upper panel. The lower panel shows the residuals to that fit. As one can see, the phase of the periodic planet signal does not match during the time frame from about 2002 to 2005, shortly after the long gap in the overall RV time series. This mismatch can also be seen in the residuals plot; during that same time frame from about 2002 to 2005 the residuals are larger than at other times.}
\end{figure*}

\begin{figure}
\centering
\includegraphics[width=9cm]{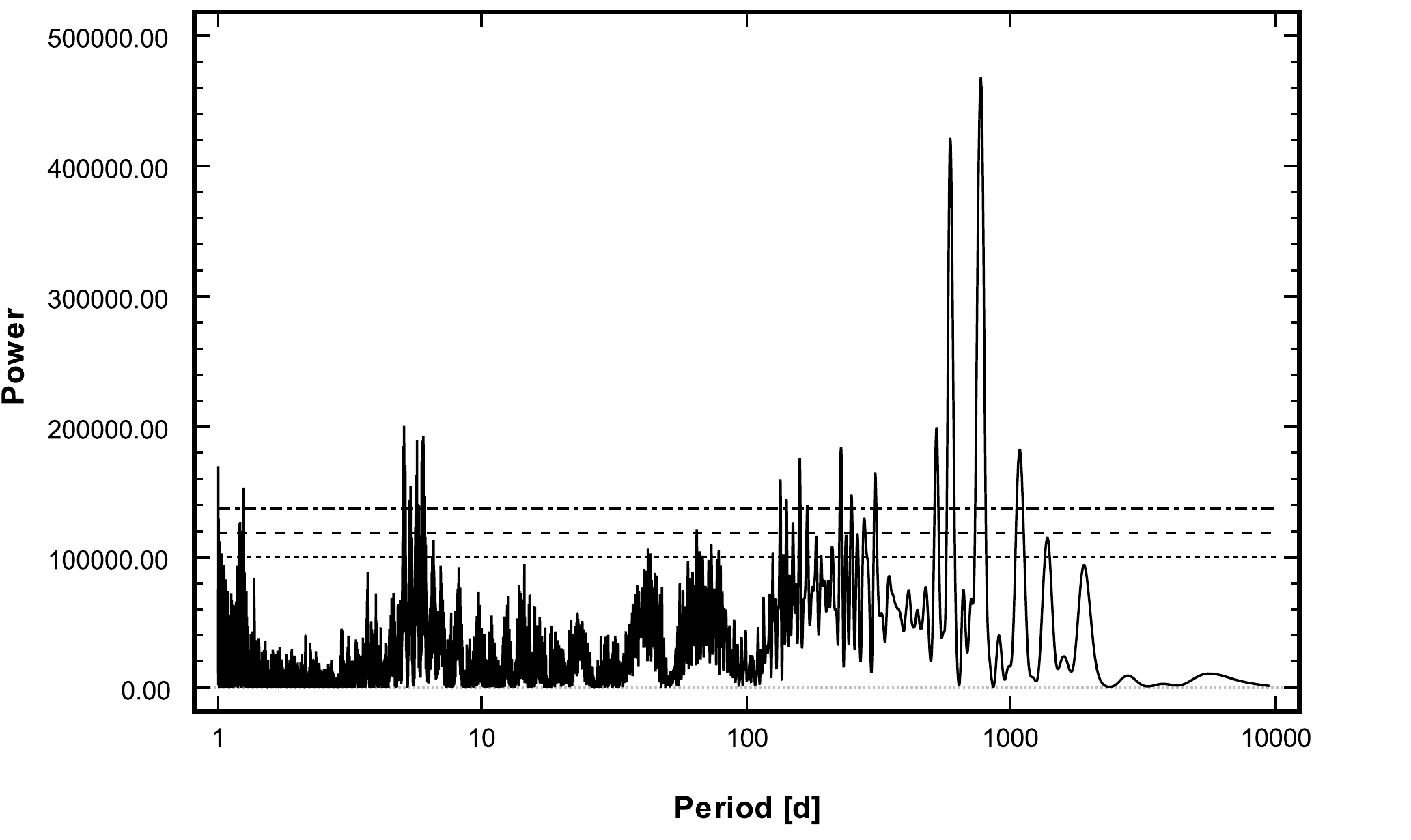}
\caption{\label{farr_periodogram} Periodogram of the residuals to the planet fit from \cite{Farr_2018}, shown in Fig.~\ref{farr_fit}. One can clearly see that there is excess power at periods around 5--6~days as well as even higher and very significant power at periods of 550--800 days, close to the period of 629 days attributed to the putative planet.}
\end{figure}

In Fig.~\ref{farr_fit} we show the full H15 data set supplemented both with the SONG data from \citet{Farr_2018} and our own Lick data presented in this paper, together with the Keplerian fit from \citet{Farr_2018} (since the values for the
periastron time and the longitude of the periastron were not given, we adjusted these parameters ourselves). The lower panel of Fig.~\ref{farr_fit} shows the residuals with respect to the Keplerian fit, and Fig.~\ref{farr_periodogram} shows the periodogram of these residuals.

The periodogram of the residuals is quite illustrative. It is clearly visible that there are several period regimes where significant power is still present in the residuals. As expected, there is power at various periods around 5 days, which corresponds to the acoustic oscillations which we did not model here. The power around periods of 5 days is mainly due to the high-cadence SONG data; these peaks in the L-S periodogram are absent in Fig.~\ref{fig:Losca_comb_trendremoved} where the SONG data have not been used. 

The largest peaks however occur at periods which are of the same order as the fitted Keplerian period of 629 days. The two strongest peaks in Fig.~\ref{farr_periodogram} are seen at periods of 586 and 766 days, above and below the assumed Keplerian period. The CARMA modeling will not be able to reduce these peaks, since they are located in a completely different period range (around 5 days, as opposed to longer than 500 days, respectively). These additional long periods are also seen in Fig.~\ref{fig:Losca_comb_trendremoved}, where the power at the acoustic periods is absent, so that the long additional peaks cannot be aliases resulting from combinations of similar shorter periods. 

However, these peaks that are of the same order as the assumed orbital period are the ones that challenge the existence of the planet. They are responsible for the observed deviations from a strictly periodic Keplerian orbit, such as an apparent phase shift in 2004 or the decrease in the L-S power of the assumed orbital period in 2006/2007, which we discussed in the main body of the paper and led us to conclude that the planet claimed to orbit Aldebaran might not exist. 

Thus, the modelling of the acoustic modulations does not help with the interpretation of the much longer periods in the RV data of Aldebaran, and it will also not enable higher sensitivity for planet detection around giant stars in general (in particular since most Lick stars are much less evolved than Aldebaran and thus have even shorter acoustic oscillation periods, while at the same time lacking high cadence SONG data). 

\begin{table}
\caption{\correction{Bootstrap uncertainties for the one-planet solution for the Lick data and for the combined RV data of Lick and H15, as well as for the two-planet solution. The uncertainties for the orbital solutions as obtained from the covariance matrix are shown in Tables~\ref{tab:orbital_sol_comparison} and \ref{tab:orbital_sol_2_planets}, respectively.}}
\label{tab:orbital_sol_comparison_bootstrap}
\centering
\begin{tabular}{l c c c l}
\hline\hline \vspace{-0.3cm} \\
\correction{Parameter} & 
\correction{Lick} & \correction{Lick + H15} & \correction{Planet 1} & \correction{Planet 2}\\
\hline \vspace{-0.3cm} \\
\correction{$\unit[P]{[days]}$} & \correction{$\pm 3.01$} & \correction{$\pm 4.96$} & \correction{$\pm 3.84$} & \correction{$\pm 4.36$} \\ 
\correction{$\unit[T_{\text{Periastron}}]{[JD]}$} & \correction{$ \pm 20.4$} & \correction{$ \pm 28.0$} & \correction{$\pm 41.9$} & \correction{$\pm 188.7$} \\
\correction{$\unit[K_1]{[ms^{-1}]}$} & \correction{$ \pm 11.9$} & \correction{$\pm 6.17$} & \correction{$\pm 6.59$} & \correction{$\pm 7.74$} \\
\correction{$e$} & \correction{$ \pm 0.07$} & \correction{$ \pm 0.06$} & \correction{$\pm 0.05$} & \correction{$\pm 0.14$} \\
\correction{$\omega$ [deg]} & \correction{$ \pm  10.6$} & \correction{$ \pm 10.4$} & \correction{$\pm 15.1$} & \correction{$\pm 113.2$} \\
\hline
\end{tabular}
\end{table}

\onecolumn
\section{Radial-velocity data from Lick Observatory for Aldebaran}
\begin{longtable}{c r@{}c@{}l r@{}c@{}l}
\caption{\label{tab:RV_data_Lick1} Radial-velocity measurements with error from Lick Observatory for Aldebaran.}\\
\hline\hline
Julian Date & RV& &[$\mathrm{ms}^{-1}$]& $\sigma_\mathrm{RV}$& &[$\mathrm{ms}^{-1}$]\\
\hline
\endfirsthead
\caption{continued.}\\
\hline\hline
Julian Date & RV& &[$\mathrm{ms}^{-1}$]& $\sigma_\mathrm{RV}$& &[$\mathrm{ms}^{-1}$]\\
\hline
\endhead
\hline
\endfoot
 2451781.02441 & $ -108$&.&$6 $ &  3&.&4 \\
 2451782.02539 & $  -40$&.&$1 $ &  3&.&1 \\
 2451783.02637 & $   10$&.&$4 $ &  3&.&5 \\
 2451784.02539 & $  -59$&.&$1 $ &  3&.&0 \\
 2451807.98047 & $ -107$&.&$3 $ &  2&.&9 \\
 2451811.94043 & $   -2$&.&$4 $ &  3&.&1 \\
 2451854.83301 & $  -78$&.&$7 $ &  3&.&4 \\
 2451856.81543 & $  -49$&.&$5 $ &  3&.&5 \\
 2451897.78516 & $  -51$&.&$1 $ &  4&.&0 \\
 2451898.80762 & $  -95$&.&$0 $ &  3&.&8 \\
 2451899.80371 & $  -77$&.&$4 $ &  3&.&5 \\
 2451901.80664 & $   63$&.&$2 $ &  4&.&2 \\
 2451929.69824 & $   67$&.&$7 $ &  3&.&6 \\
 2451930.75098 & $  101$&.&$5 $ &  3&.&4 \\
 2451931.68750 & $   83$&.&$3 $ &  3&.&8 \\
 2451932.72168 & $   81$&.&$2 $ &  3&.&3 \\
 2452165.00293 & $   79$&.&$1 $ &  3&.&3 \\
 2452165.97168 & $  126$&.&$2 $ &  3&.&3 \\
 2452167.02734 & $  117$&.&$2 $ &  3&.&2 \\
 2452174.99609 & $  221$&.&$6 $ &  3&.&5 \\
 2452175.93262 & $  233$&.&$8 $ &  4&.&2 \\
 2452176.93262 & $   71$&.&$0 $ &  3&.&2 \\
 2452192.94434 & $  111$&.&$8 $ &  3&.&4 \\
 2452193.87012 & $  140$&.&$2 $ &  4&.&1 \\
 2452194.93164 & $   63$&.&$6 $ &  4&.&0 \\
 2452205.87988 & $   59$&.&$9 $ &  4&.&2 \\
 2452206.87793 & $   46$&.&$3 $ &  4&.&1 \\
 2452207.89648 & $  137$&.&$9 $ &  3&.&5 \\
 2452222.82715 & $   14$&.&$5 $ &  3&.&8 \\
 2452223.84668 & $  121$&.&$5 $ &  3&.&5 \\
 2452258.70410 & $   22$&.&$1 $ &  3&.&8 \\
 2452259.78516 & $  -62$&.&$5 $ &  4&.&0 \\
 2452295.69238 & $   50$&.&$9 $ &  4&.&3 \\
 2452297.69824 & $  -44$&.&$9 $ &  4&.&8 \\
 2452307.66211 & $   62$&.&$0 $ &  3&.&6 \\
 2452362.66895 & $ -109$&.&$2 $ &  4&.&1 \\
 2452363.65527 & $ -112$&.&$0 $ &  4&.&5 \\
 2452495.02539 & $  -73$&.&$2 $ &  3&.&3 \\
 2452496.01465 & $  -98$&.&$5 $ &  3&.&7 \\
 2452497.02539 & $  -56$&.&$2 $ &  3&.&8 \\
 2452506.01758 & $ -161$&.&$8 $ &  3&.&7 \\
 2452506.97559 & $  -88$&.&$7 $ &  3&.&6 \\
 2452507.97656 & $  -46$&.&$5 $ &  3&.&7 \\
 2452517.98438 & $ -234$&.&$7 $ &  3&.&9 \\
 2452520.03027 & $  -67$&.&$7 $ &  3&.&4 \\
 2452528.97559 & $ -208$&.&$3 $ &  2&.&9 \\
 2452530.01367 & $ -192$&.&$2 $ &  3&.&5 \\
 2452530.93945 & $ -163$&.&$7 $ &  3&.&1 \\
 2452531.97949 & $ -173$&.&$3 $ &  3&.&0 \\
 2452532.93262 & $ -241$&.&$8 $ &  4&.&7 \\
 2452541.96289 & $ -119$&.&$1 $ &  3&.&3 \\
 2452542.93262 & $ -168$&.&$4 $ &  4&.&3 \\
 2452543.93359 & $ -249$&.&$1 $ &  3&.&4 \\
 2452559.98438 & $ -179$&.&$7 $ &  5&.&1 \\
 2452560.88574 & $ -164$&.&$4 $ &  4&.&9 \\
 2452561.95898 & $ -195$&.&$8 $ &  4&.&4 \\
 2452571.89258 & $ -178$&.&$9 $ &  3&.&9 \\
 2452573.89746 & $  -83$&.&$3 $ &  4&.&0 \\
 2452589.87988 & $  -74$&.&$8 $ &  3&.&9 \\
 2452590.86328 & $  -55$&.&$2 $ &  3&.&9 \\
 2452603.84277 & $  -85$&.&$8 $ &  5&.&2 \\
 2452604.75391 & $  -56$&.&$9 $ &  4&.&3 \\
 2452605.76562 & $  -67$&.&$7 $ &  4&.&1 \\
 2452615.78027 & $  -29$&.&$3 $ &  4&.&3 \\
 2452616.80371 & $  -68$&.&$6 $ &  3&.&6 \\
 2452617.75293 & $  -64$&.&$1 $ &  5&.&1 \\
 2452664.71875 & $  -97$&.&$2 $ &  4&.&5 \\
 2452667.70703 & $   10$&.&$7 $ &  3&.&5 \\
 2452668.68848 & $    8$&.&$8 $ &  3&.&6 \\
 2452699.62305 & $   63$&.&$0 $ &  5&.&5 \\
 2452720.64941 & $   96$&.&$8 $ &  4&.&1 \\
 2452879.99707 & $   89$&.&$5 $ &  3&.&1 \\
 2452881.99805 & $   95$&.&$6 $ &  3&.&8 \\
 2452898.99121 & $  171$&.&$5 $ &  3&.&2 \\
 2452899.94727 & $  118$&.&$9 $ &  3&.&6 \\
 2452900.97266 & $   93$&.&$1 $ &  3&.&7 \\
 2452932.88477 & $  -11$&.&$2 $ &  4&.&1 \\
 2452933.88086 & $   37$&.&$6 $ &  3&.&5 \\
 2452934.88477 & $   53$&.&$3 $ &  3&.&5 \\
 2452963.82031 & $  138$&.&$1 $ &  4&.&5 \\
 2452964.77148 & $   69$&.&$7 $ &  4&.&5 \\
 2452965.88184 & $  -56$&.&$1 $ &  4&.&8 \\
 2452966.81641 & $  -59$&.&$0 $ &  3&.&9 \\
 2452985.84570 & $   86$&.&$2 $ &  5&.&6 \\
 2453022.68555 & $ -166$&.&$1 $ &  4&.&9 \\
 2453089.63477 & $   45$&.&$7 $ &  5&.&2 \\
 2453232.02637 & $  -20$&.&$0 $ &  3&.&2 \\
 2453233.01367 & $   -1$&.&$3 $ &  3&.&6 \\
 2453234.02051 & $  -12$&.&$2 $ &  3&.&4 \\
 2453236.03516 & $  -96$&.&$6 $ &  3&.&3 \\
 2453266.96484 & $   24$&.&$6 $ &  3&.&8 \\
 2453269.02051 & $  101$&.&$3 $ &  3&.&2 \\
 2453270.98340 & $  106$&.&$2 $ &  3&.&3 \\
 2453287.00781 & $  119$&.&$7 $ &  3&.&9 \\
 2453324.82154 & $   13$&.&$6 $ &  4&.&1 \\
 2453354.73512 & $   51$&.&$1 $ &  3&.&7 \\
 2453358.81419 & $    6$&.&$8 $ &  3&.&5 \\
 2453400.72502 & $  203$&.&$1 $ &  3&.&8 \\
 2453424.65026 & $   79$&.&$1 $ &  3&.&9 \\
 2453442.64420 & $   78$&.&$7 $ &  3&.&8 \\
 2453444.62922 & $   93$&.&$9 $ &  4&.&8 \\
 2453446.62799 & $  100$&.&$7 $ &  4&.&0 \\
 2453613.01655 & $  -49$&.&$8 $ &  3&.&5 \\
 2453614.00594 & $  -31$&.&$1 $ &  3&.&7 \\
 2453701.85940 & $ -120$&.&$6 $ &  3&.&9 \\
 2453741.72650 & $ -108$&.&$3 $ &  4&.&1 \\
 2453791.69205 & $   26$&.&$1 $ &  4&.&2 \\
 2453803.62553 & $ -126$&.&$2 $ &  8&.&8 \\
 2453970.03250 & $  125$&.&$0 $ &  3&.&4 \\
 2453981.97465 & $  170$&.&$2 $ &  3&.&0 \\
 2453983.96214 & $   75$&.&$5 $ &  3&.&2 \\
 2453985.98241 & $  135$&.&$6 $ &  3&.&2 \\
 2454058.85436 & $  -67$&.&$0 $ &  4&.&0 \\
 2454082.79825 & $ -116$&.&$2 $ &  3&.&7 \\
 2454182.62638 & $ -127$&.&$4 $ &  4&.&4 \\
 2454345.99395 & $ -101$&.&$2 $ &  3&.&3 \\
 2454348.01564 & $ -135$&.&$8 $ &  3&.&6 \\
 2454419.86277 & $  -22$&.&$2 $ &  3&.&5 \\
 2454420.71007 & $  -36$&.&$9 $ &  3&.&8 \\
 2454421.93501 & $   64$&.&$6 $ &  3&.&9 \\
 2454440.90300 & $  112$&.&$8 $ &  3&.&6 \\
 2454444.78801 & $  107$&.&$6 $ &  4&.&4 \\
 2454481.76068 & $  243$&.&$9 $ &  3&.&8 \\
 2454485.78462 & $  174$&.&$5 $ &  3&.&6 \\
 2454502.78558 & $  249$&.&$9 $ &  3&.&4 \\
 2454504.71505 & $  263$&.&$2 $ &  3&.&6 \\
 2454505.67145 & $  265$&.&$7 $ &  3&.&8 \\
 2454507.64262 & $  214$&.&$9 $ &  3&.&4 \\
 2454557.64373 & $  241$&.&$3 $ &  3&.&6 \\
 2454711.99325 & $ -107$&.&$5 $ &  3&.&3 \\
 2454713.97765 & $ -201$&.&$4 $ &  3&.&3 \\
 2454714.97050 & $ -183$&.&$4 $ &  3&.&3 \\
 2454715.95207 & $ -119$&.&$5 $ &  3&.&4 \\
 2454716.97995 & $  -99$&.&$0 $ &  3&.&0 \\
 2454754.84463 & $ -104$&.&$0 $ &  3&.&5 \\
 2454755.96156 & $  -92$&.&$9 $ &  4&.&0 \\
 2454756.90991 & $ -164$&.&$9 $ &  2&.&7 \\
 2454778.91292 & $ -128$&.&$4 $ &  3&.&4 \\
 2454809.78796 & $ -124$&.&$9 $ &  4&.&7 \\
 2454883.76091 & $ -141$&.&$8 $ &  4&.&7 \\
 2455065.04123 & $  185$&.&$5 $ &  4&.&8 \\
 2455066.04274 & $  325$&.&$5 $ &  4&.&3 \\
 2455098.01046 & $  254$&.&$9 $ &  4&.&5 \\
 2455120.99838 & $  216$&.&$4 $ &  5&.&5 \\
 2455121.92263 & $  179$&.&$1 $ &  5&.&6 \\
 2455155.89098 & $  216$&.&$1 $ &  6&.&9 \\
 2455278.65458 & $  164$&.&$8 $ &  4&.&7 \\
 2455421.02537 & $ -214$&.&$2 $ &  3&.&7 \\
 2455450.98654 & $  -99$&.&$5 $ &  3&.&5 \\
 2455463.95773 & $ -175$&.&$6 $ &  4&.&2 \\
 2455466.94491 & $  -67$&.&$2 $ &  4&.&0 \\
 2455519.81435 & $ -180$&.&$5 $ &  4&.&4 \\
 2455566.80296 & $ -118$&.&$6 $ &  4&.&6 \\
 2455589.70358 & $ -159$&.&$6 $ &  3&.&3 \\
 2455590.72405 & $ -138$&.&$5 $ &  3&.&4 \\
 2455593.67863 & $ -134$&.&$1 $ &  3&.&9 \\
 2455620.71306 & $   13$&.&$3 $ &  4&.&5 \\
 2455651.67484 & $  -59$&.&$4 $ &  3&.&9 \\
 2455804.99527 & $  121$&.&$8 $ &  3&.&7 \\
 2455807.01751 & $  125$&.&$2 $ &  3&.&6 \\
 2455829.04349 & $  163$&.&$6 $ &  3&.&4 \\
 2455829.94395 & $  161$&.&$1 $ &  3&.&6 \\
 2455832.94313 & $  227$&.&$5 $ &  4&.&0 \\
 2455865.97124 & $   86$&.&$5 $ &  4&.&3 \\
 2455895.91321 & $   82$&.&$2 $ &  6&.&8 \\
\hline
\hline
\end{longtable}

\end{appendix}

\end{document}